\documentclass[prl, twocolumn]{revtex4-2}
\usepackage{amssymb}
\usepackage{amsmath}
\usepackage{epsfig}
\usepackage{bm}
\usepackage{xcolor}

\begin{document}

\title{ Hydrodynamics of two-dimensional electrons
 due to scattering by disorder }

\author{
D. R. Raskulov, K. A. Baryshnikov, and P. S.  Alekseev
}

\affiliation{  Ioffe Institute,    Politekhnicheskaya 26,   Saint~Petersburg  194021,  Russia }

\begin{abstract}

The hydrodynamic regime of electron transport, induced
 by fast inter-electron collisions, was discovered 
 in  high-quality nanostructures in recent ten years.
However, signs of hydrodynamic transport, primarily, the giant negative
magnetoresistance, were  observed even at very low temperatures,
when electron-electron scattering is too weak to affect the transport.
 To address this puzzle, here we develop a theory of {\em mixed},
 hydrodynamic and non-Markovian, magnetotransport 
  of two-dimensional  electrons
at zero temperature in  samples with weak but still important disorder.  Namely, we
account for {\em both} the memory effects at  electron scattering
by  localized defects in magnetic field and  an unconventional viscosity effect
due to electron scattering by  defects in bulk  and by rough sample edges.  Solution of the model yields
a strong negative magnetoresistance,  which exhibits at zero magnetic field a sharp maximum
 in narrower  samples or a blunt maximum in wider samples. 
 This and other our results explain various properties of  the giant negative magnetoresistance
observed on   ultra-high-quality GaAs quantum wells, thereby we apparently reveal
the nature of   low-temperature magnetotransport in these systems.

\end{abstract}

\maketitle

{\em 1. Introduction. } Hydrodynamic electron transport 
can be implemented in conductors with very low  densities  
of defects~\cite{Gurzhi_rev,1n,3n,4n}.
Usually, formation of viscous electron  flows
 is induced  by frequent inter-particle  collisions together
  with the scattering of electrons on the rough or curved sample edges.
 The hydrodynamic regime was first reliably
identified in ultra-pure samples of layered palladium
cobaltate~\cite{Moll},  in single-layered
graphene~\cite{graphene_1,Polini_Geim,Levitov_Falkovich},
 and in GaAs quantum wells~\cite{je_visc,Gusev_1,recentest_,
 exps_neg_1,exps_neg_2,exps_neg_3,exps_neg_4,Gurzhi_Shevchenko,Keser,f1}.
In the latter case, it was first detected
 by the strong temperature-dependent negative magnetoresistance
in classical magnetic fields~\cite{je_visc,Gusev_1,recentest_,exps_neg_1,exps_neg_2,
exps_neg_3,exps_neg_4,Gurzhi_Shevchenko} and, then,
by the dependence of the sample resistance on its complex
 geometry~\cite{Keser}.

In recent ten years many bright effects of the hydrodynamic
 electron transport  were studied experimentally and theoretically.
Essentially two-dimensional (2D) flows of the electron fluid   in samples with
 macroscopic obstacles  were examined~\cite{d1,d1_new,d2,d2_new,disks,IVG_DGP_}.
In Refs.~\cite{Scaffidi2017,a,Holder,Nature,Nature2,Gusev2}  the transition
from the hydrodynamic to the ballistic  flow regime with an increase
of magnetic field  were investigated for ultra pure samples.
 Other stationary and low-frequency hydrodynamic transport effects
were considered in many works, in particular, in
 Refs.~\cite{Afanasiev2022,Denisov2022,c,Glazov,Zohrabyan,Zohrabyan2,Denisov2023,Alekseev_2023,
Afanasiev_at_al_2025,Alekseev_Dmitriev,Polini,Levin2024,Alekseev_Semina_2025}.
Unexpected effects of high-frequency hydrodynamic transport
 were theoretically studied
in Refs.~\cite{vis_res_0,vis_res_1,vis_res_2,Semiconductors,Afanasiev_2023,new},
and, possibly, were observed  in best-quality samples
in experiments~\cite{Smet,Dai2010,Hatke2011,Bandurin2022}.

Despite of the discovery of many hydrodynamic effects in
 the 2D electron fluid and their successful explanation within
 the  conventional hydrodynamics of viscous flows formed due to
inter-particle  collisions, there are several problems and open questions
unresolved  up to now.
First of all, it looks absolutely mysterious within this picture, why
 the hydrodynamic-like transport effects, such as the giant negative
magnetoresistance~\cite{Gurzhi_Shevchenko,je_visc,Gusev_1,recentest_}
and the viscoelastic resonance at the doubled cyclotron
frequency~\cite{vis_res_1,vis_res_2,Afanasiev_at_al_2025}, are
observed in high-quality samples even at the lowest temperatures~\cite{Gusev_1,
recentest_,exps_neg_1,exps_neg_2,exps_neg_3,exps_neg_4,Dai2010,Hatke2011},
 when  the inter-particle scattering is too weak
 and the ballistic or  Ohmic regimes are to  be realized~\cite{f,Maxim}.
So a  big question arises: what are the mechanisms of magnetotransport at zero temperature  in  record-quality structures?

A strong negative magnetoresistance at~zero temperature can be induced
by the  memory  effects due to the scattering
of 2D electrons  on localized defects or/and  smooth disorder in magnetic field~\cite{m1,m2,m3,m11,m10,m4,m6,m7,m9}.
These effects are induced by the occurrence of electron trajectories
 of different types with respect to scattering  on such disorder 
 due to the cyclotron rotation.
As a result, the electron dynamics becomes non-Markovian and correlated in time, 
in particular, the standard kinetic equation is not applicable.
 This leads to   a strong negative
magnetoresistance~\cite{m1,m2,m3,m11,m10,m4,m6,m7} and to
microwave-induced resistance oscillations~\cite{m12,m8}.
 However, the observed shape and properties   of the low-temperature  negative 
 magnetoresistance~\cite{f0} differ essentially from
the ones  predicted by these theories
(see discussion in~\cite{exps_neg_4}).

In this  way, new ideas are required to resolve
this puzzle in magnetotransport  of high-quality
conductors.

Here we propose  a  model of low-temperature
magnetotransport  of 2D electrons that takes into account 
 {\em both}
(i)~the   strong memory effects  in electron scattering by localized defects in magnetic field
 and (ii)~the hydrodynamic effects induced by  the unconventional viscosity
due to electron scattering by  defects in bulk and by rough sample edges. 
 We construct and solve the balance transport equations for electron  flows
 in a long sample with   the strong rare localized defects and an additional  weak
disorder between them. The resulting magnetoresistance
contain the contributions  from~the viscosity and  from~the memory effects,
which has different properties and can manifest themselves together in a given sample.
 At zero magnetic field such magnetoresistance
exhibits   a sharp peak in the narrow samples,
where the hydrodynamic contribution,
dominates or  a blunt maximum in wider samples, where  the Ohmic memory-induced
contribution becomes main. We calculate the magnetoresistance
for realistic parameters of  disorder in high-quality GaAs quantum wells, and
 also  account the effects from electron-electron and electron-phonon  scattering
 in order to describe the evolution of  magnetoresistance  with temperature~$T$.
 We compare  the  results with experimental data and demonstrate that
our theory describes the  giant negative magnetoresistance
observed in various samples of  high-quality GaAs quantum wells  very well.  
Thereby  we apparently  provide  a 
solution  of the problem of the nature of magnetotransport
 in  these  structures at~$T \to 0$.

{\em 2. Model and balance equations.}  An analysis of experimental data on resistivity of ultra-pure GaAs/AlGaAs quantum wells
in zero  magnetic field at lowest temperatures was performed  
 in~Ref.~\cite{Huang_Shklovskii_Zudov}. Its results are cited
and analyzed in Supplemental Material (SM)~\cite{SM}.
Here  we  use the conclusion of~\cite{Huang_Shklovskii_Zudov}
  that the scattering of 2D electrons by the screened charged (Coulomb) defects
inside the 2D layer often dominates in the low-temperature transport in these systems.
 We also discuss in SM that such defects and other-type weaker defects
can be approximately treated as  localized defects with some effective radius $r_0$
 and density $n_d \sim d^{-2}$ (herewith $r_0 \ll d \ll W$, where $d$ is the mean  distance between the  defects and $W$ is the sample  width),  and a weak
 unspecified disorder, ``dust'',  between the localized defects, see Fig.~1.
The scattering lengths on the localized defects,~$l_1 = v_F \tau_1$, and
on the ``dust'',~$l_1 ^d=v_F \tau_1^d$, differ substantially,~$\tau_1 \ll \tau_1^d $,
but  even the length $l_1$ can be sufficiently large, comparable with
the  sample width~$W$   (here $v_F$ is the Fermi velocity).

In a perpendicular magnetic field electrons move along circular cyclotron orbits
of the radius $R_c = v_F /\omega_c$ with the   period $T_c = 2 \pi /\omega_c $
[here $\omega_c = eB/(mc)$ is the cyclotron frequency
and $m$ is the electron  mass].
When $R_c \lesssim l_1$, a certain fraction of electrons
are not scattered by the  localized defects [see Fig.~1($b$)]; they are named
the circling (c-)  electrons~\cite{m1,m2,m4}.  In the absence of the  ``dust'',
they stay indefinitely long on  the collisionless cyclotron trajectories.
The remaining electrons are scattered by localized defects, herewith they can scatter
 by different defects [the ``traveling''  (t-) electrons,
see Fig.~1]  or scatter by same defects several times, forming ``rosette'' trajectories
(not shown in Fig.~1).  As it was demonstrated in~\cite{m2,m3}, the role of the latter
is relatively minor for steady-state transport.

In the presence of a weak ``dust'' between the localized defects, scattering
of the c-electrons occurs during the characteristic time~$\tau_1^d $,
which can lead to a strong shift of the cyclotron orbit and
a subsequent collision of an electron with a  localized defect,
 that is changes of the electron type slowly occurs, $c \leftrightarrow t$.
Nevertheless, due to the  assumption $\tau_1^d \gg  \tau_1 $, the relative fraction of
 the c-electrons  is approximately given by the probability:
\begin{equation}
\label{P}
   P \,  = \, \exp(  \, -\, T_c\, /\,\tau_0\,)
   \:, 
\end{equation}
of passing a cyclotron circle  without scattering on the localized defects
(here~$\tau_0$ is the departure time due to the scattering  by localized defects;
$\tau_0 $ and~$ \tau_1 = l_1 / v_F $ differ by a numeric factor~\cite{m2}).

\begin{figure}[t!]
\centerline{\includegraphics[width=.77 \linewidth]{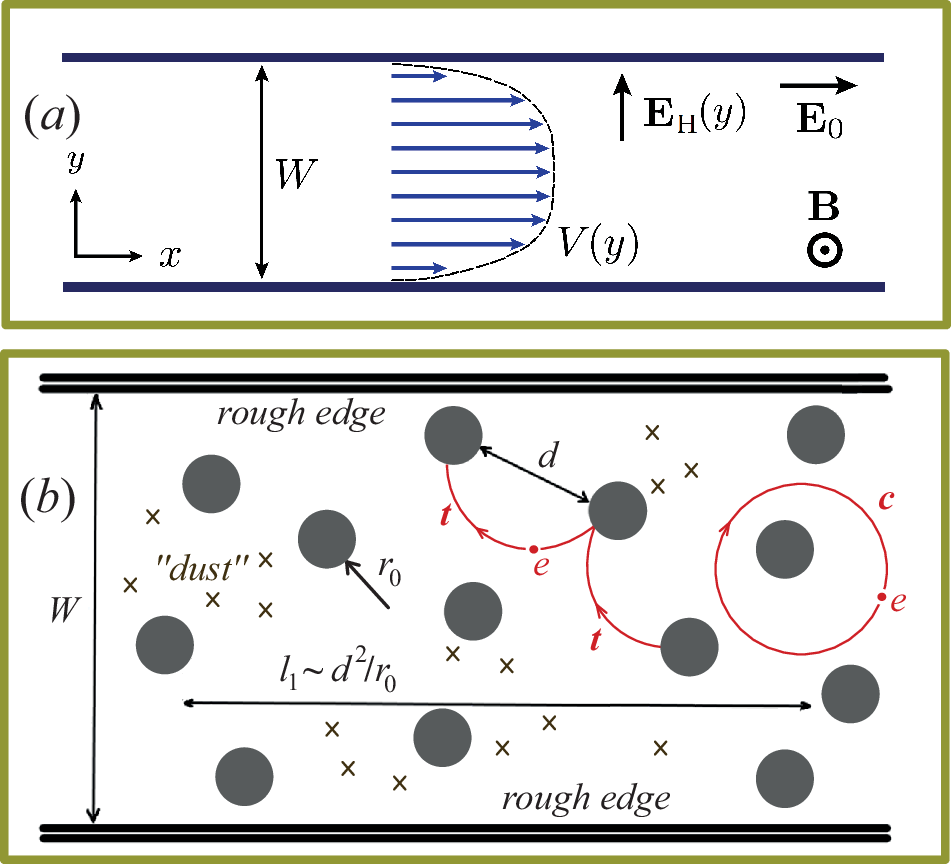}}
\caption{
($a$):
Long sample with non-interacting 2D electrons forming an Ohmic-hydrodynamic
 flow due to scattering on  defects
and rough longitudinal edges.  The  perpendicular magnetic
 field~$ \mathbf{B} $, the external electric field~$\mathbf{E}_0$
due to applied voltage, and the internal Hall electric field~$\mathbf{E}_H(y)$
are shown, as well as the hydrodynamic  velocity
profile~$V_x( y) \equiv V(y)$.
($b$):
 Sample with localized defects and
  weak unspecified defects between them (``dust'').
The two types of 2D electron trajectories appear in  magnetic field:
the c-electrons, which do not collide with defects, and
 the t-electrons, which scatter by  different defects.
}
\end{figure}

Now, based on the approaches of~\cite{m8,m9,Alekseev_2023,eh,long},
we formulate the phenomenological balance  transport equations for   t- and c-electrons~\cite{f3,f4}.

We denote all the values for the two components of the system, the t- and
c-electrons,  by the superscripts~$\alpha =t,\,c$. Accordingly, $\delta n^\alpha(\mathbf{r},t)$
are the inequilibrium perturbations of their equilibrium densities,~$ n_0 (1-P) $
and~$ n_0 P $;  $\mathbf{q}^\alpha (\mathbf{r},t)\, $  
are their 2D flows, which are related to their
hydrodynamic  velocities  as:~$\mathbf{q}^t (\mathbf{r},t) \, = \, (1-P) \, n_0 
 \mathbf{V}^t (\mathbf{r},t)  $  and~ $\mathbf{q}^c  (\mathbf{r},t) \, = \, P \, n_0 
\mathbf{V}^c (\mathbf{r},t)  $. The total electron flow  is:~$\mathbf{ q} = \mathbf{q}^t
+\mathbf{q}^c $,   the perturbation of the electron density is $ \delta n  =
\delta n ^ t + \delta n ^ c $,  and the current density  is~$\mathbf{ j} = e\mathbf{ q} $.

First of all, we formulate   the balance equations for the densities
of  t- and c-electrons, accounting the transitions between the two groups
of 2D electrons,~$t \leftrightarrow c$:
\begin{equation}
 \label{bal_eq_dens}
 \partial \, \delta n^\alpha / \partial t
 \, +\,
 \mathrm{div} \, \mathbf{q}^\alpha
 \, =\,
 -\,
 \Gamma _ \alpha \delta   n^\alpha
 \, +\,
 \Gamma _ { \bar{\alpha} } \delta  n^{ \bar{\alpha} }
 \:,
\end{equation}
where we denote $ \bar{\alpha }= c, \, t $ for $ \alpha = t, \, c $.
The rates~$ \Gamma _ \alpha $ of the transitions $t \leftrightarrow c$ are primary
due to the scattering on the ``dust''.  It is natural to consider that  such transition are not affected by the memory effects, thus $ \Gamma _ \alpha \equiv  \Gamma $.   For a long   sample
we have $ \mathrm{div} \, \mathbf{q}^\alpha = \partial q_y^\alpha / \partial y$ 
 (see Fig.~1). 
Equation~(\ref{bal_eq_dens}) 
is written in the way  that ensures
conservation of  the total number
 of electrons. Namely, the boundary condition of the closed edges,
$q _y | _{y = \pm W/2 } = 0  $,  and integration by time of~(\ref{bal_eq_dens})
in the absence of extra electrons  at $t=0$, $\delta n  | _{t=0 } = 0  $,
 leads to the equality $ \int _{-W/2} ^{W/2} dy \: \delta n  (y,t) =0$
 at any~$t$.  Next, provided there is no additional $c \leftrightarrow t$ transitions
in the very  vicinities of the edges, 
 we should impose the zero boundary conditions
on the each flow component:  $ q_y^\alpha | _{y = \pm W/2 } = 0  $.

The electric field  in the Hall bar sample has
 the form: $\mathbf{E}(y)\,=\,(\,E_x\,,\,E_y (y)\,)$,
where $E_x \equiv E_0$ is the applied field and  $E_y (y) \equiv E_H  (y) $
is the Hall electric field  appearing due to perturbation of the charge density,
$e \delta n(y)$ (in this paragraph we omit the possible dependencies on time
for brevity).  For example, if we consider a structure with
 a bulk metallic gate, located at the small distance~$d_g$
near the 2D layer,   the so-called ``gradual channel approximation'' yields:
$ E_H (y) = ( 4 \pi d_g e / \kappa )  \, \partial \, \delta n / \partial y $
(here $\kappa$ is the dielectric constant of the media between  the layer and the gate).
In  structures without a gate, the Hall field  is a non-local operator
of the density perturbation,~$E_H [\delta n(y') ](y)$.
 Perturbations of the densities~$\delta n^\alpha$  also induce the  partial pressures
 $ \delta P ^ \alpha (y)  = (m v_F^2 /2) \, \delta n ^ \alpha  (y) $.
The hydrostatic forces, $ - \nabla \delta P ^ \alpha (y) $, together with
the electric Hall force, $eE_H(y)$, compensate the magnetic Lorentz forces,
$(e/c)[ \mathbf{q}^\alpha (y) \times \mathbf{B} ]$,
acting on the t- and c-components.

In this way, the balance equation for  t-electrons, accounting the magnetic,
electric and  hydrostatic forces,  the viscosity effect and momentum relaxation
in the bulk,  in a Hall bar sample is (see Fig.~1):
\begin{equation}
\label{bal_eq_flows_of_t-el}
\begin{array}{c}
 \displaystyle
  \frac{\partial \mathbf{q} ^t }{\partial t }
   =
    \frac{n_0 e }{m} \, (1-P)   \, \mathbf{E} (y,t)
   -
    \frac{1}{m}  \, \nabla   \delta P^t
    +
    \\
    \displaystyle
    +
    \frac{n_0}{m} \, (1-P) \, \frac{\partial  \boldsymbol{\Pi}^t }{\partial y }
    +
    \omega_c \,  [ \mathbf{q} ^ t  \times \mathbf{e}_z]
    -
    \frac{\mathbf{q} ^t}{ \tau_1 }
    \:,
 \end{array}
\end{equation}
where   $\boldsymbol{\Pi}^t \, = \,  (\,  \Pi_{xy}^t \, , \,  \Pi_{yy}^t \, )$
is the viscous part of the momentum flux carried by the t-electrons, per one electron
and~$ \nabla =  \mathbf{e}_y \, \partial / \partial y  $  for the considered geometry.
For simplicity, let us consider that the longitudinal sample edges are very rough,
corresponding to the sticking boundary conditions:~$ q_x^t|_{y=\pm W/2} = 0 $.

The motion equation for $\boldsymbol{\Pi}^t$ has its usual form, with the diffusive, 
magnetic and relaxation terms,
similar to the equation for one-component systems of 2D electrons with
inter-particle collisions (see,  for example, Ref.~\cite{je_visc}):
\begin{equation}
\label{bal_eq_mom_flux_of_t-el}
  \frac{\partial  \boldsymbol{\Pi}^t }{\partial t }
  =
  -
   \frac{mv_F^2}{4}  \frac{\partial  \mathbf{V}^t  }{\partial y }
   +
   2\omega_c \,  [\boldsymbol{\Pi}^t \times \mathbf{e}_z]
   -
   \frac{\boldsymbol{\Pi} ^t }{\tau_2}
\:,
\end{equation}

Dynamics of c-electrons is mainly controlled   by the collisionless motion
in the fields~$\mathbf{E}$ and~$\mathbf{B}$  and weak scattering on dust,
with the rates~$1/\tau_{1,2}^d \ll 1/\tau_1$. Owing to this relation,
 the momentum flux of c-electrons, $n_0P\boldsymbol{\Pi}^c$, 
turns out  negligible~\cite{ff}
and their balance equation is:
\begin{equation}
 \label{bal_eq_flows_of_c-el}
\begin{array}{c}
\displaystyle
  \frac{\partial \mathbf{q} ^c }{\partial t }
   =
     \frac{n_0 e }{m}  P \,  \mathbf{E} (y,t)
   -
  \frac{ \nabla     \delta P^c}{m}
    +
    \omega_c \,  [ \mathbf{q} ^ c  \times \mathbf{e}_z]
    -
    \frac{\mathbf{q} ^t}{ \tau_1^{d} }
     .
 \end{array}
\end{equation}

{\em 3. Stationary flows in relatively narrow samples. } Let us consider
magnetotransport in this  system in  the  stationary regime.
First of all,   from  Eqs.~(\ref{bal_eq_dens}) 
 at~$\Gamma _\alpha \equiv  \Gamma $  and from
 the boundary conditions $q_y^\alpha | _{y = \pm W/2 } = 0  $
we obtain the conservation of the total number of each type electrons:
 $\int _{-W/2} ^{W/2} dy \: \delta n ^ \alpha (y) =0$.
 Next, an analysis, being  analogous to the one in Refs.~\cite{therm-el,long}
 for electron-hole systems shows, that 
solutions of equations~(\ref{bal_eq_dens})-(\ref{bal_eq_flows_of_c-el})
contain the two characteristic lengths: 
 the Gurzhi length,  $l_G (B)=\sqrt{\eta_{xx} \tau_1} = \sqrt{l_1l_2 /(1+ \tilde{\beta}^2 )}/2$, where $l_2= v_F\tau_2$ is the shear stress relaxation length, and
the   characteristic length of the ``$t  \leftrightarrow c $'' 
  transitions,~$ l_{tc } =  \sqrt{D_0/[ \Gamma  \, (1+\beta^2 ) ]} $.
  If a sample is not too wide, $W \ll l_{tc } $ or even  $W \lesssim l_{tc } $,     
  the last processes do not affect the transport substantially. Namely, 
we should put  $\Gamma  \to 0 $ in this limit,  
then equations~(\ref{bal_eq_dens})
 and conditions~$q_y^\alpha | _{y = \pm W/2 } = 0  $
yield~$ q_y^\alpha (y) \equiv 0 $. In other words, the flows
 are directed only along the sample: $\mathbf{q} ^\alpha (y)
 = q^\alpha (y)  \, \mathbf{e}_x $. In the opposite limit,  $W \gg l_{tc}$,
 the density perturbations  $\delta n ^ \alpha $ and the ``$t  \leftrightarrow c $'' 
  transitions are substantial 
  in the near-edge regions with the widths~$l_{tc}$,
  while in the bulk region  of 
  the sample non-zero flows $ q_y^\alpha (y) $ appear.

For the regime $W \lesssim l_{tc } $, we  obtain
from Eq.~(\ref{bal_eq_mom_flux_of_t-el})
 the expressions for the momentum flux components
  of t-electrons:~$\Pi_{xy}^t = - m\, \eta_{xx}  ( V_{x}^t)' $
and~$\Pi_{yy} ^t =  m\, \eta_{xy}  ( V_{x}^t )' $,
where the  prime denotes the derivative~$d/dy$,
$\eta_{xx} = (v_F ^2 \tau_2/4) / (1+ \tilde{\beta} ^2 ) $
and~$\eta_{xy} = \tilde{\beta} \, \eta_{xx}$ are the shear diagonal
 and Hall viscosities, and  $\tilde{\beta} = 2 \omega_c \tau_2 $. 
These formulas indicate that   shear stress  relaxation on the defects in bulk, provided strong electron momentum relaxation at  rough edges, produce inhomogeneous electron flows and momentum flux, which gives  an unconventional   viscosity.

Equation~(\ref{bal_eq_flows_of_t-el}) for the t-electrons flows
takes the form:
\begin{equation}
 \label{eq_q_t-el__stationary}
    \begin{array}{l}
    (1-P) \, \sigma_0 \,E_0
   \, + \,
    \tau_1   \eta_{xx} (q_x^t)''
    \,-\,
    q_x^t
    \,=\,
    0 \:,
    \\
   ( 1-P)\,  \sigma_0  \, E_H(y)
      -
    D_0  \,  (   \delta n^t )'
    \,-
    \\
 \quad \quad   \quad \quad  \quad   \;\;
      \: \, - \, \tau_1   \eta_{xy} (q_x^t)''
    \,-\,
    \beta \, q_x^t
    \, = \,
    0  \:,
    \end{array}
\end{equation}
where  $ \sigma_0 = n_0 e \tau_1 / m $, $D_0 = v_F^2 \tau_1 /2$
is the diffusion coefficient at~$B=0$, and $\beta =\omega_c \tau _ 1 $.
  As  the viscosity term appears in Eq.~(\ref{eq_q_t-el__stationary}),
 boundary conditions on $q_x^t(y)$, for example,  $ q_x^t|_{y = \pm  W/2 } =0 $,  
 are needed to solve it.

Equation~(\ref{bal_eq_flows_of_c-el}) for the flows of c-electrons in 
a stationary case for  a Hall sample becomes as follows:
\begin{equation}
\label{eq_q_c-el__stationary}
    \begin{array}{l}
    P  \,\sigma_0 \, E_0
     \, - \,  ( \tau_1/\tau_1^d)  \,q_x^c
    \,=\,
    0
    \:,
    \\
     P\,   \sigma_0   \,E_H(y)
     \,-\,
    D_0 \,(   \delta n^c )'
   \,-\,
    \beta  \, q_x^c
    \,=\,
    0\:.
    \end{array}
\end{equation}
Unlike Eq.~(\ref{eq_q_t-el__stationary}), equation~(\ref{eq_q_c-el__stationary}) is
Ohmic-like, that is  only the momentum relaxation at scattering on the ``dust'' in bulk
is accounted,  thus there is no  need to impose any boundary conditions on~$q_x^c(y)$.

From the $x$-components of Eqs.~(\ref{eq_q_t-el__stationary})
  and~(\ref{eq_q_c-el__stationary}) we immediately obtain the results
for the flows of c- and t-electrons:
\begin{equation}
\label{q}
  \left.
  \begin{array}{c}
    q_x^t(y)
      \\
     q_x^c(y)
\end{array}
\right\}
=
\frac{en_0E_0}{m}
\left\{
\begin{array}{c}
\displaystyle    
(1-P)\, \tau_1 \Big [ 1- \frac{\cosh(y/l_G) }{ \cosh(\xi) } \Big]
     \\ 
      P\, \tau_1^d
\end{array}
\right.
,
\end{equation}
where  
$\xi = W/(2l_G) $ is the parameter characterizing the relative importance of the Ohmic and
 the hydrodynamic contributions. The total electric current  is:
 $I= e \int_{-W/2} ^{W/2} [q_x^t(y)+q_x^c(y)]  \, dy  $.
 The averaged sample resistivity is defined as:~$\varrho_{xx} = E_0/ \langle j \rangle $, 
where $\langle j \rangle  = I/W$.  From Eq.~(\ref{q}) we obtain
the hydrodynamic-memory-induced magnetoresistance, 
 being the main result of our theory:
\begin{equation}
\label{res_I}
 \varrho_{xx}=\varrho_D 
  \, /\, 
 \big[ \,
   ( \tau_1^d/ \tau_1) \, P
    \, + \,
 \big( \,
 1 \, - \,
  \tanh \xi \, / \,  \xi \,
 \big) \,(1-P) \,
 \big]
 \,,
\end{equation}
where $\varrho _D =  m/(n_0e^2\tau_1)$ is the Drude resistivity.

In~SM we also present and discuss the results for the averaged Hall
resistivity~$\varrho _{xy}= - \int_{-W/2} ^{W/2} E_H(y) \, dy   /I $,
 following  from Eqs.~(\ref{eq_q_t-el__stationary})-(\ref{q}).

Note that equation (\ref{res_I}) yields a previously unknown type
of the memory-induced Ohmic magnetoresistance for moderately wide samples, 
$ l_G \ll W \ll l_{tc}$:
\begin{equation}
 \label{MR_ohm_mem_narr}
    \varrho_{xx}^{\rm n} (B)
    \, = \,
    \varrho_ D \,/\, [  \,  (\tau_1^d/ \tau_1 ) \, P  \, + \,     (1-P)  \,]
\:,
\end{equation}
In the   limit of very wide sample, $W \gg l_{tc}$,
solution of Eqs.~(\ref{bal_eq_dens})-(\ref{bal_eq_flows_of_c-el}),
 should lead  to the well-known memory-induced bulk
Ohmic  magnetoresistance~\cite{m1,m4}:
\begin{equation}
\label{BEM}
    \varrho_{xx}^{\rm w} \, \approx \, (1-P) \: \varrho_D
     \:.
\end{equation}
It implies
an important role of the density perturbation $\delta n^\alpha $
and the transitions $c \leftrightarrow t $ in the near edge regions
 with the widths $\sim l_{tc}$, similarly as it takes place
  for magnetotransport 
in electron-hole systems~\cite{eh,therm-el,long}.
 Narrow-sample  magnetoresistance~$ \varrho_{xx}^{\rm n} (B)$~(\ref{MR_ohm_mem_narr})
is significantly stronger than the wide-sample one, $ \varrho_{xx}^{\rm w} (B)$:
 the first one falls  at  $\beta \gtrsim  2 \pi$,
as~$ \varrho_{xx}^{\rm n}  \sim \varrho_D / [ ( \tau_1^d /  \tau_1 ) \beta ] $,
  in contrast to the second one, which falls  
as~$ \varrho_{xx}^{\rm w} \sim \varrho_D / \beta $.

{\em 4. Results and comparison with experiments. } In Fig.~2($a$) we plot
 the mean resistivity~$\varrho_{xx}$~(\ref{res_I})   as a function of 
 the dimensionless magnetic field,   $ \beta = \omega_c \tau _ 1  $, 
together with: (i)~the result~$  \varrho_{xx}^{\rm w} $~(\ref{BEM})
  for the memory-effect-induced magnetoresistance in very wide
samples, $W \gg l_{tc}$;   and (ii)~the result
 of Refs.~\cite{Gurzhi_rev,Gurzhi_Shevchenko,je_visc},
 $ \varrho_{xx}^{\rm G} \approx \varrho_D / [1-\tanh \xi/\xi]$,
 for the  hydrodynamic-Ohmic Gurzhi  flow without   memory effects.
We see from Fig.~2($a$)  that the ``mixed'' hydrodynamic-memory  
magnetoresistance~$\varrho_{xx}(B)$~(\ref{res_I})
inherits  both the properties of~$ \varrho_{xx}^{\rm G } (B)$ 
(the narrow peak at $\beta \lesssim 1$),
as well as of~$ \varrho_{xx}^{\rm n}$~(\ref{MR_ohm_mem_narr}) 
 (the wide blunt maximum at $\beta \lesssim 2\pi$
  and the relatively fast decrease at $\beta \gtrsim 2\pi$
 with saturation to the value~$ \varrho_ D / (\tau_1^d/ \tau_1 ) $).
We also see that the magnetoresistances~$\varrho_{xx}$~(\ref{res_I})
and~$ \varrho_{xx}^{\rm n} $~(\ref{MR_ohm_mem_narr})
in the region~$\beta \gtrsim 2\pi$
is actually much stronger ,
 than the wide-sample one~$ \varrho_{xx}^{\rm w}$.

\begin{figure}[t!]
\centerline{\includegraphics[width=.99 \linewidth]{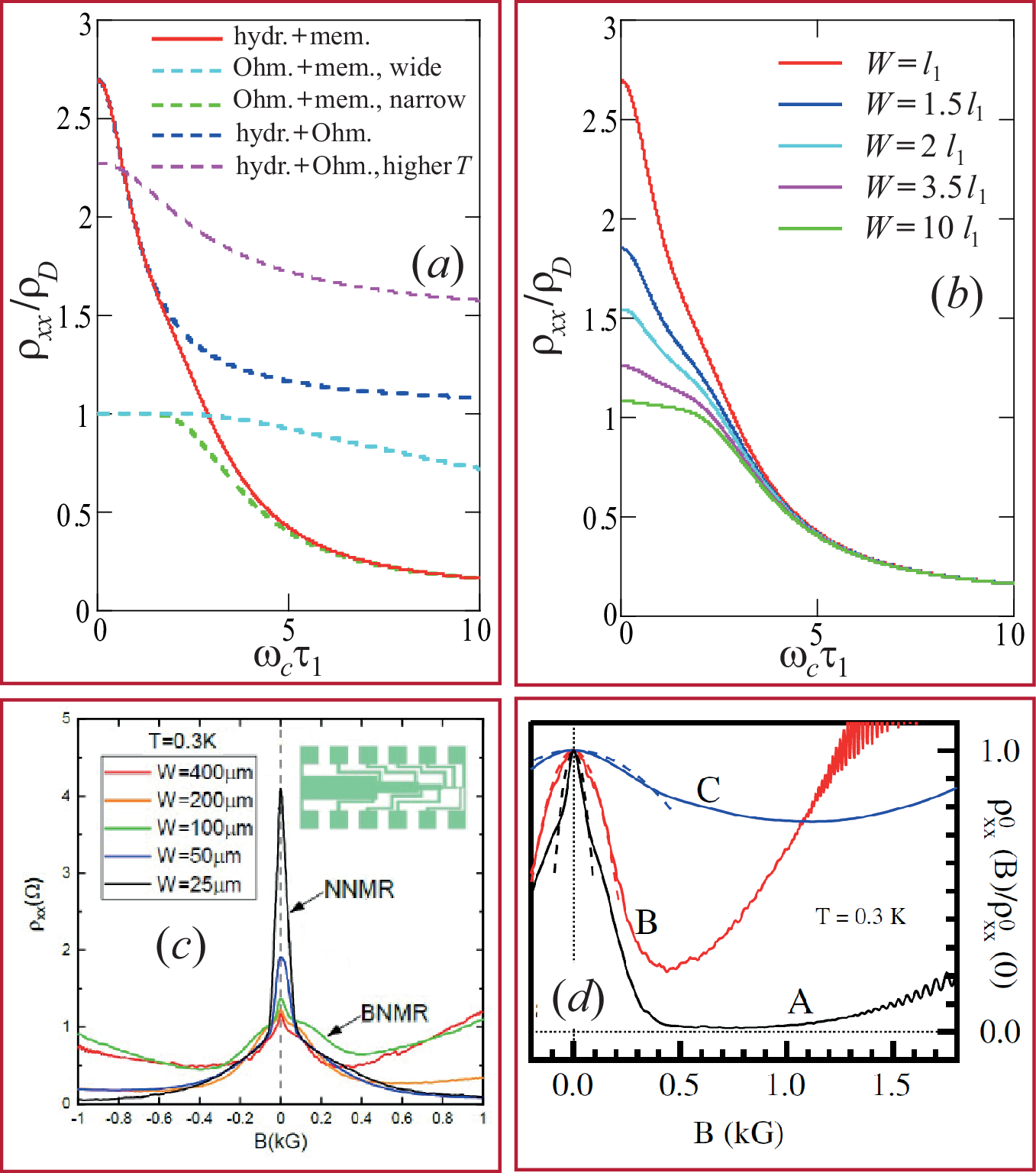}}
\caption{
($a$):
Mean sample resistivity~$\varrho_{xx}$ as a function of~$\beta = \omega_c \tau_1$
obtained  for:
(i) the current hydrodynamic-memory model [Eq.~(\ref{res_I})],
(ii) the Gurzhi hydrodynamic-Ohmic model [Eq.~(\ref{res_I}) at~$P=0$],
 (ii) the Ohmic-memory model for a wide, $W \gg l_{tc}$, [Eq.~(\ref{BEM})]
 and a narrower, $W \lesssim l_{tc}$, [Eq.~(\ref{MR_ohm_mem_narr})] samples.
 For all the curves except the violet one 
  the parameters:  $W= l_1 $, $ l_2 = 0.5 \, l_1 $, $l_0  =  0.5 \, l_1 $,
and~$l_1^d  =  20\, l_1 $ were used; for the violet curve~$l_1 ' = 0.7 l_1$
 and~$l_2 ' = 0.4 l_2 $, those correspond to a higher temperature.
($b$):
Resistivity~$ \varrho_{xx} $ as a function of~$\beta = \omega_c \tau_1$
at various sample widths~$W$ for  the current  model.
The lengths $ l_2 $, $ l_0 $, and $l_1^d$   are  the same as for
% the most of  curves in
panel~($a$).
($c$):
Magnetoresistance
  for the high-quality GaAs quantum well sample (drawn
inside panel) with the sections of different widths~$W$,
  figure is cited from Ref.~\cite{recentest_}.
Are observed: (i) the ``narrow negative magnetoresistance'' (NNMR)
 for  smaller~$W$, with a  narrow big peak and a wider
  blunt main part,   (ii) the ``bell-shaped
negative magnetoresistance'' (BNMR) for  larger~$W$
with a wide   blunt main part and a very small  peak.
($d$):
Magnetoresistance  GaAs quantum well of  sample~A, B, and~C
of different quality,
  measured in Ref.~\cite{Dai2010}.
    For the purest sample~A magnetoresistance is very strong and can be divided on
a narrow sharp  peak near~$B=0$ and a wider part at larger~$B$.
    Such two-part shape is similar to the BNMR in panel~($c$).
For the dirtier samples~B and~C
  magnetoresistance becomes weaker, wider and blunter.
  }
\end{figure}

In Fig.~2($b$) we present the resistivity within our model~$\varrho_{xx}(\beta)$
for different sample widths.  We see that for moderately wide samples,
when
 $ l_{1,2} \ll W \ll l_{tc}$ (we remind that $l_G(B) <\sqrt{l_1 l_2 }/2 $), the Ohmic-memory contribution~$\varrho_{xx}^{\rm n}$~(\ref{MR_ohm_mem_narr})
dominates at any~$B$.  With the decrease of~$W$ the blunt maximum at $B =0 $
becomes sharper, triangular-like.  In narrow samples,  $W \sim l_{1,2}$,
a sharp large peak in~$\varrho_{xx}(\beta)$  appears  due to the viscosity-induced contribution
in the  small~$B$, where  $\tilde{\beta} \lesssim 1$.
The decrease of the predicted magnetoresistance~(\ref{res_I}),
  $ r= \varrho_{xx}|_{ \beta = 0 } / \varrho_{xx}|_{ \beta \gg 1} $ is estimated
as~$ \, \sim\tau_1^d / \tau_1 \gg 1 $, which is in a stark contrast with the hydrodynamic-Ohmic Gurzhi 
magnetoresistance without memory, $\varrho_{xx}^{\rm G}$,
 for which~$r \sim 1$ at~$W \gtrsim l_2$ [see Fig.~2($a$)].

Quite diverse forms of the giant negative
magnetoresistance were observed   in various experiments
[see Figs.~2($c$,$d$) and Fig.~S5 in SM].
Its  common feature  in high-quality samples
 is the more or less large amplitude: the ratio~$r$ can reach the values $20-50$.
 In many samples magnetoresistance curves  consist of two distinct regions:
 a narrow peak in the region of relatively low magnetic fields and a wide main part
in higher magnetic fields [see Figs.~2($c$,$d$)], 
 while in some samples the curve is single-component
  (see~SM).   As we have seen above, our theory
also predicts  a large magnetoresistance magnitude~$r \gg 1 $
and, depending on the parameters,
  a single-component or two-part shape of the curves~$\varrho_{xx}(\beta)$, 
  with  a narrow peak
at  $\tilde{\beta } \lesssim 1$  due to the  viscosity and a wide blunt main part
due to the  memory effects.   As for a quantitative comparison, 
the experimental magnitude of~$\varrho_{xx} (B)$ and
 the characteristic  magnetic field~$B^*$  where a strong drop of~$\varrho_{xx} (B)$  
 occurs  rather well correspond predictions of our theory 
 for realistic sample parameters  (see~SM). 
 % It should be noted that the shape and parameters of the defects in the samples under discussion 
 % are not precisely defined, so it is impossible to achieve a literal superposition of the theoretical and experimental curves.

Next, we extend our consideration to  the magnetotransport
at non-zero temperatures. We limit ourselves
to expanding the developed model in the direction of 
taking into account that all the scattering processes and
the rates $1/\tau _{1,0,2}$ and $1/\tau _1^d$  acquire contributions 
from electron-electron and  electron-phonon collisions with the increase of~$T$.
 Herewith  we neglect the increase of  the  rate~$ \Gamma    $ of 
 the $t \leftrightarrow c $ transitions with~$T$,
 for which we implied above the limit 
 $  W \ll l_{tc} \sim  \Gamma   ^{-1/2} $ and, thus,  continue 
 to consider the flows of t- and c-electrons as independent.
We use the known results for the electron-phonon and electron-electron
 rates for  the model  of an almost ideal Fermi gas
interacting with acoustic phonons  (for details see SM).
Then we substitute the total scattering  rates  into Eq.~(\ref{res_I})
and obtain the resistance~$\varrho_{xx}(B,T)$ as a function of magnetic field and temperature.
In SM we present the obtained results and compare them with experimental data. 
 Here we only
mention that   both in theory and in experiments:
(i) the absolute value of the resistance at $B=0$ can decrease with~$T$  (this is the Gurzhi effect, being  the famous fingerprint of electron hydrodynamics),
while (ii) the negative magnetoresistance  of all types broadens and/or
disappears with the increase of~$T$  more or less fast.

{\em 5. Conclusion and acknowledgments. }  We have developed  a theory of the low-temperature semi-hydrodynamic magnetotransport of 2D electrons.
Our model  is based on the simultaneous accounting  of 
the unconventional 
viscosity effect,  caused by the electron scattering by defects 
 in the sample bulk and the strong scattering
at the rough sample edges, as well as the memory effects
 due to the correlated in time electron dynamics
in magnetic field in the presence of the localized defects. 
The  theory yields a strong negative
 magnetoresistance, non-trivially depending on the sample parameters, which  explains very well the experimental data on the giant
negative magnetoresistance  in various high-quality  GaAs quantum well samples. The obtained results apparently  resolve the problem of the origin
 of the low-temperature hydrodynamic-like 
 magnetotransport of 2D~electrons in  such high-quality structures.

We thank M.~I.~Dyakonov for discussions of experimental data 
on magnetotransport in the high-quality  GaAs quantum wells
as well as  for drawing our attention to the problem of a nature of
 the giant negative magnetoresistance in the low-temperature limit in these systems,  which is addressed here.
We thank Y.~M.~Beltukov,  M.~M.~Glazov, L.~E.~Golub, 
 and B.~I.~Shklovskii  for  fruitful discussions of this~work.

This work  was carried out under the state assignment
of the Ministry of Science and Higher Education of the
Russian Federation.

\clearpage

\setcounter{equation}{0}
\setcounter{figure}{0}

\renewcommand{\thefigure}{S\arabic{figure}}
\renewcommand{\thesection}{S\Roman{section}}
\renewcommand{\theequation}{S\arabic{equation}}

\onecolumngrid
\begin{center}

{\Large  {\bf Supplemental material   to the manuscript
 ``Strange hydrodynamics of two-dimensional electrons at zero temperature'' }
\linebreak
}

{
\large D. R. Raskulov, K. A. Baryshnikov, and P. S.  Alekseev
\linebreak \linebreak}
{\small
 Ioffe  Institute, Politekhnicheskaya 26,
  194021,   St.~Petersburg,   Russia
\linebreak
}
\end{center}

{\small
Here we present the details of our theoretical model
of the hydrodynamic/non-Markovian magnetotransport
  of 2D electrons in samples with localized defects as well as
 the results obtained within this model for the Hall effect
 and for  temperature dependencies of magnetoresistance.
 We also review here experimental data on the magnetotransport
 in various high-quality GaAs quantum well and graphene structures
 and discuss  further possibly developments of the proposed model.
\linebreak
\linebreak
\linebreak}
\twocolumngrid

\section{1. Defects in ultra-high-quality GaAs  quantum wells}
In work~\cite{Huang_Shklovskii_Zudov}  an analysis of experimental data
on the resistance of ultra-pure GaAs/AlGaAs quantum well samples with
different 2D electron concentrations~$n_0$ (in the absence of a magnetic field)
was performed.    Were taken into account the scattering of electrons
on various types of defects: the screened charged (Coulomb) impurities
in and near the 2D layer;
 spatial fluctuations of the quantum well width and of the composition~$x$
of the Al$_x$Ga$_{1-x}$As solution (``interface roughnesses'' and ``alloy disorder'');
 the remote defects far from the 2D layer.
  The analysis of Ref.~\cite{Huang_Shklovskii_Zudov} 
  shows that the screened Coulomb impurities in the 2D layer play a dominant role
in the scattering of 2D electrons at sufficiently
small   electron densities~$n_0 \lesssim 2 \cdot 10^{11}$~cm$^{-2}$,
while for the structures with high 2D electron densities~$n_0 $,
a more important role in transport
is played by scattering on the interface roughnesses.
However, the performed analysis also admits that both these two
 and other contributions
to the electron scattering can be comparable in some intervals of~$n_0$.

Herewith the defect  densities (or, saying mo generally,
the disorder strengths) are typically extremely  low for
 the best-quality GaAs quantum well samples,
namely, the observed low-temperature resistivities~$\varrho_0$ and electron mobilities~$\mu_0$
corresponds to the electron mean free paths~$l_1$ of the order of 100~$\mu$m.
 In samples of lower quality than analyzed in Ref.~\cite{Huang_Shklovskii_Zudov},
  but in which the giant negative magnetoresistance is still well observed,
  the scattering lengths are smaller in an order of magnitude,
  that is, are of the order of  10~$\mu$m.
These values are comparable to the sample sizes~$W$ or the space scales~$W_{eff}$
 of macroscopic inhomogeneities  in a sample controlling the effective width
of conductive channels. Such inhomogeneities   can arise, for example,
 due to  the frequently occurring macroscopic ``oval'' defects
with a radius of~10-20~$\mu$m and a typical distances between
 them about~50~$\mu$m~\cite{d1_new}.

\begin{figure}[t!]
\centerline{\includegraphics[width=.99 \linewidth]{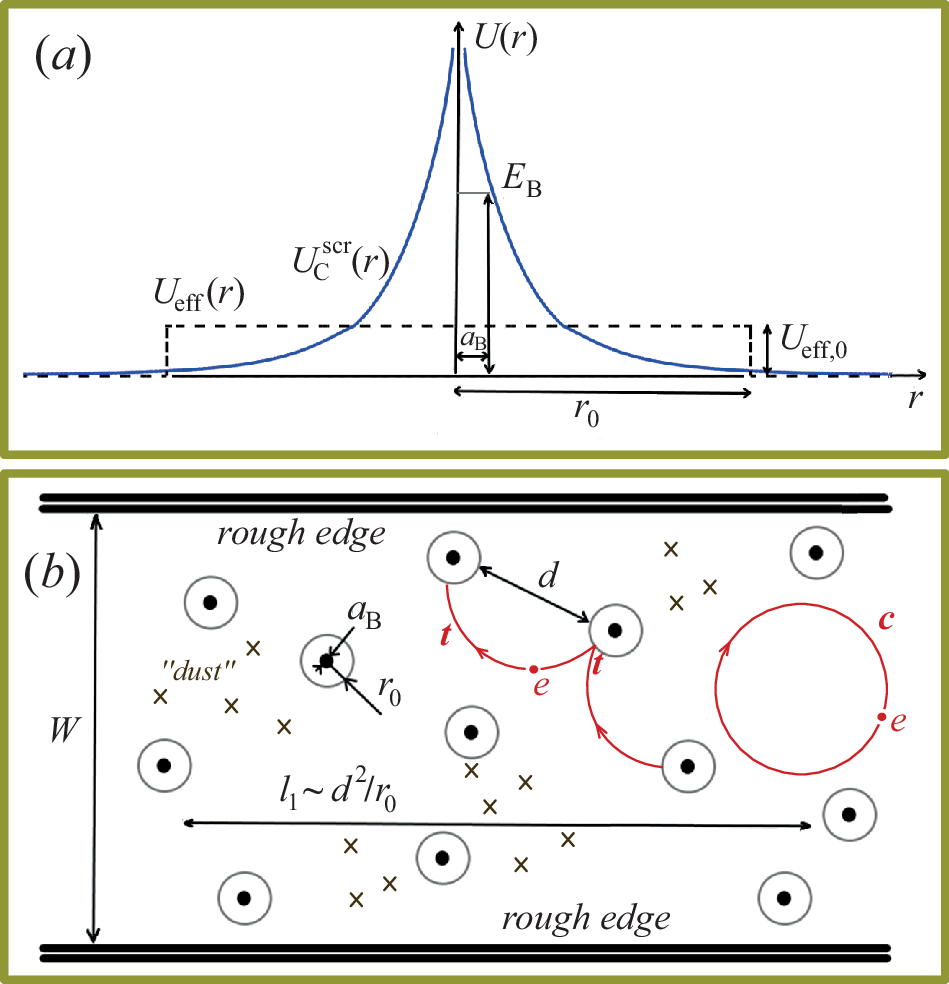}}
\caption{
($a$):
  Electric potential~$U_C^{scr}(r)$ of a Coulomb defect,
 screened by 2D electrons,
 and the effective localized potential~$U_{eff}(r)$,
 on which the exact potential~$U_C(r)$
 may be changed in order to simplify
the consideration of the memory effects in magnetic field.
($b$):
 Sample with localized defects, the screened Coulomb defects, and
  weak unspecified defects between them (``dust'').
The two types of 2D electron trajectories are shown in  magnetic field:
the c-electrons, which do not collide with defects, and
 the t-electrons, which scatter by  different defects.  
Electrons on ``rosette'' trajectories, which collide
 multiple times with the same defect, are not shown.
 It is supposed that~$W \gg d \gg r_0 $, $l_1 \lesssim W$,
 while the relations between  $R_c$~and $l_1$, $W$, $d$  can be arbitrary.
  }
\end{figure}

In this way, the main examples of the systems, for which we have developed
the theory  of  hydrodynamic-like magnetotransport in this work,
are such  ultra-pure GaAs quantum wells with the screened Coulomb defects and 
 with other important disorders, first of all, interface roughnesses.
 It is very important for our  model  that the first defects are localized.

  In accordance with the analysis of Ref.~\cite{Huang_Shklovskii_Zudov},
the predicted here magnetotransport regime is expected to be better pronounced
  in the samples with lower electron
densities,~$n_0 \lesssim  2\cdot 10^{11}$~cm$^{-2}$,
where the scattering by  the defects of the first type  dominates.
Apparently, the decrease of the magnetoresistance
amplitude~$\varrho_{xx}(0)/\varrho_{xx}(\infty)$
with  the increase of~$n_0$ in some samples support this conclusion, see Fig.~2
in Ref.~\cite{exps_neg_3}.
 However, the considered here regime  seems to be well realized also
  in samples with higher 2D electron densities
  and, possibly, relatively weaker role of the localized defects (see next sections below).  This  issue about the relation between $n_0$ and the importance of the localized defects requires further analysis.

The size of  the ``nucleus'' of a screened Coulomb defect is on the order of
 the Bohr radius;   $a_B = 10$~nm in GaAs [see Fig.~S1($a$)].
At $r \sim a_B$ the magnitude of the screened Coulomb potential
   is of the order of the Bohr energy~$E_B$.
 Beyond this radius, the screened Coulomb
 potential decays quite rapidly,  inversely proportional
 to the distance to the defect,~$r$,
in the third power, $U_C^{scr}(r) \sim 1/r^3$ at $ r \gg a_B $~\cite{Rytova}.
We assume that such defects can be approximately considered
as sparsely distributed over a sample,
approximately localized defects with
some {\em effective } radius $ r_0 > a_B $ and
the {\em effective } height~$U_{eff,0} < E_B$ [see Fig.~S1($a$)],
corresponding, at the given density of these defects   $N_C \sim d^{-2}$,
to the observed total electron mean free path $l_{tr}  \approx   l_1  $
in the limit $T \to 0$ [see Fig.~S1($b$)].
 The momentum relaxation length~$l_1 $ determined
by the scattering by the localized defects
is given by the well-known formula  $l_1 \sim d^2/r_0$~\cite{m2}.

Outside the radius $r_0$, a random relatively weak potential (``the potential
of dust'') exists,  being caused both by the ``tails'' of the screened Coulomb
potential  from the charged defects~$ U_C(r) \sim 1/r^3 $, $r>r_0$, as well as by all
other types of defects. An important part of our model is the assumption that
the potential of the ``dust'' is weak,
having a long scattering length, $ l_1^d \gg l_1 $.
Note that such length~$l_1^d = v_F \tau_1^d$,
can be  much larger than the sample size,~$l_1^d \gg W$ or ~$l_1^d \gg W_{eff}$,
but this is not a problem
 as this lengths does not enter directly  in the transport characteristics,
but only the ration~$l_1^d/l_1$ will be the  relevant value in the theory.

In this way, we use in the current work such very rough model of the two-component disorder in the considered
structures, consisting of: (i) the localized defects with the effective radius~$r_0$ and
the density~$N_C \sim d^{-2}$, leading to large scattering lengths,~$ l_1 \sim d^2/r_0$,
of the order of the sample width~$W$;
 and (ii) much weaker delocalized disorder (``dust'') between the localized defects
 with the scattering rate~$1/\tau_1^d \ll 1/\tau_1$.
\\
\\
\section{2. Hall effect }
The theory  developed  in the main text allows to find the distribution of
the c- and t-electron inequilibrium densities, $\delta n ^ \alpha (y)$,
and the corresponding  Hall electric field~$ E_H (y) $. These values are calculated from
the electrostatic equation for $E_H[\delta n]$ and
from the second lines of
 Eqs.~(\ref{eq_q_t-el__stationary}) and~(\ref{eq_q_c-el__stationary}), after substitution
 in them the formulas~(\ref{q}) for the flows~$q_x^{t,c}(y)$.
 Note that both the  hydrostatic and the electrostatic 
 forces~$-\nabla P ^ \alpha  $   and~$ e E_H (y)   $
 are important  for  the Hall effect in the considered two-component system of c- and t-electrons.
Below we describe the main points and  the result of such consideration
 for a mathematically simpler case of a gated structure,
where the functional $E_H[\delta n(y')](y)$ is local by~$y$.

\begin{figure}[t!]
    \centering
    \includegraphics[width=1.0\linewidth]{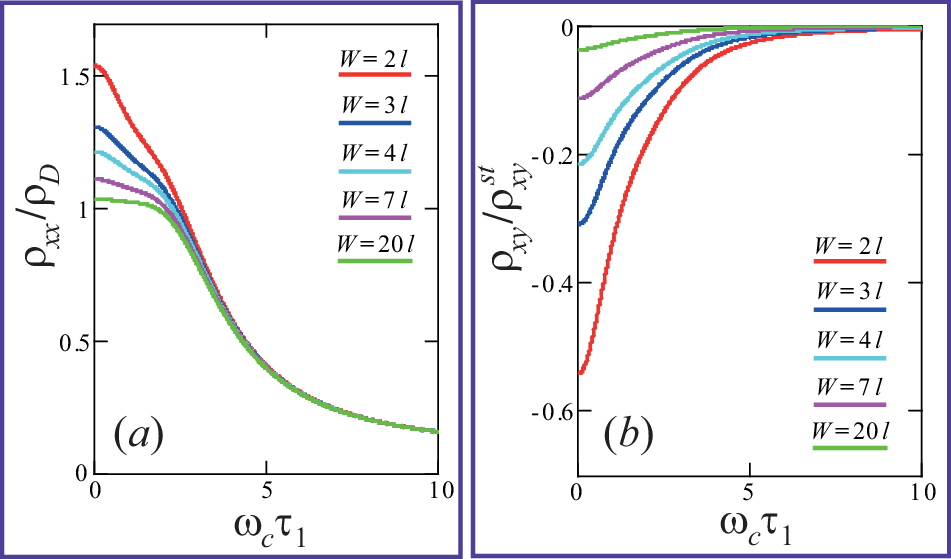}
    \caption{
    Mean sample longitudinal~$\varrho_{xx}$~($a$) and
     the Hall~~$\varrho_{xy}$~($b$) resistivities as  functions of
the dimensionless magnetic field~$\beta = \omega_c \tau_1$
in the units of
the Drude resistivity~$\varrho_D = m / (n_0 e^2 \tau_1 ) $ and
the standard Hall resistivity~$\varrho_{xy}^{st} = B/(n_0ec)  $, respectively.
The calculations were performed
for the parameters:   $ l_2 = 0.5 \, l_1 $, $l_0  =  0.5 \, l_1 $,
 $l_1^d  =  20\, l_1 $ and various sample widths $W= (2,3,4,7,20)\:l_1 $.
 }
    \label{pic:relaxation_lengths}
\end{figure}

In the $y$-components of Eqs.~(\ref{eq_q_t-el__stationary}) and~(\ref{eq_q_c-el__stationary}),
 the   forces,~$ e E_H  
 \sim - d (\delta n ^t + \delta n ^c )/dy $
and~$-d \, \delta P^{t,c} /dy \sim -d \delta n^{t,c}/dy $ 
 balance the  magnetic Lorentz forces
in the $y$-direction, $\sim \beta\, q _x^{t,c}$,
 for each t- and c-components of the electron system.
However, usually the magnitude of the Hall terms~$eE_H $
 is much larger than  the magnitudes of the hydrostatic terms~$-\nabla \delta P^\alpha$
 in two-component 2D systems of general type (without special symmetries
 of the two components, which do present only in the electron-hole systems
 at the charge neutrality point, see Refs.~\cite{eh,therm-el,long}).
Such  relation:
\begin{equation}
 \label{app}
  |\nabla \delta P^\alpha| \, \ll \, |eE_H | \,  n_0
\:, 
\end{equation}
is guaranteed by the inequality $s \gg v_F$, leading to the inequality~$s^2 \tau_1 \gg D_0 $~\cite{vis_res_2,Alekseev_2023}. Here 
\begin{equation}
s = \sqrt{4\pi n_0 e^2 d_g /(m\kappa)}
\end{equation}
is the plasmon speed
in the  gated 2D system.

We have calculated  the perturbations of the densities
of the two components,~$ \delta n ^t $ and~$ \delta n ^c $, 
and the corresponding
Hall field $E_H(y) $
and the Hall voltage~$ U_H = - \int _{-W/2} ^{W/2} dy \, E_H(y) $
in a general case of any relation between 
the terms $\nabla \delta P^\alpha$ and $eE_H $
(that is, between $s $ and $v_F$), but below we present
 only the final result for the main contribution
 to the Hall effect  in the above approximation~(\ref{app}),
  corresponding to~$s \gg v_F$.
 The mean sample Hall resistivity,
$ \varrho_{xy} = U_H/I $ in this  case  can be written in the form:
$ \varrho_{xy} = \varrho_{xy} ^{st} + \delta \varrho_{xy} $,
where
\begin{equation}
  \varrho_{xy} ^{st} \,  =  \, B/(n_0ec)
\end{equation}
  is the standard Hall resistivity
corresponding just to the balance of the electric and magnetic Lorentz  forces
in simple one-component systems, and the small non-trivial correction~$\delta \varrho_{xy} $
has the form:
\begin{equation}
\label{res_Hall}
 \frac{ \delta \varrho_{xy} }{\varrho_{xy} ^{st}}
 =
 -
 \frac{2\tau_2}{\tau_1}
 \frac{ (1-P) (\tanh \xi /\xi) }
 {
 P\, (\tau_1^d /\tau_1 )
 +
 (1-P) (1- \tanh \xi /\xi)
 }
  \,.
\end{equation}
In the absence of the memory effects ($P \equiv 0$), result~(\ref{res_Hall})
turns into the results for the Hall resistance obtained in Ref.~\cite{Scaffidi2017,c}
for the Ohmic-hydrodynamic flow.

In Fig.~S2  we draw the relative correction to the mean Hall resistivity,
$\delta \varrho_{xy} /\varrho_{xy} ^{st}$, as a function of dimensionless magnetic field,
 $ \beta = \omega_c \tau_1 $
together the relative  mean diagonal resistivity, $\varrho_{xx} /\varrho_D$~(\ref{res_I}),
 for the comparable parameters values of all the characteristic lengths
 except the length~$l_1^d$:
 $l_1 \sim l_2 \sim l_0 \sim W \ll l_1^d$.
It is seen that the correction $\delta \varrho_{xy} $~(\ref{res_Hall}) 
is always negative,
as for the usual hydrodynamic regime without memory effects~\cite{Scaffidi2017,c},
and saturates to zero with the increase of magnetic field.
As it is naturally to expect, it disappears with the increase
of the sample width, when the transport becomes Ohmic
with the memory effects [compare Figs.~S2($a$) and~($b$)].
 Calculations show that  both the functions~$\delta \varrho_{xx} (\beta)$
 and~$\delta \varrho_{xy} (\beta)$  strongly and non-trivially  depend on particular values  
 of the ratios~$l_{0,2}/l_1$ and~$W/l_1$.

It is important that in the almost hydrodynamic regime,
when $\xi \ll 1$, as well as in  the almost Ohmic regime, when $\xi \gg 1$,
the correction $\delta \varrho_{xy} /\varrho_{xy} ^{st}$~(\ref{res_Hall})
 is small and proportional to the ratios of microscopic lengths $l_1,\,l_2,\,R_c$
to the   sample width~$W$. Strictly speaking, any hydrodynamic-like theory based
 on balance equation for particle flows (in particular, the current theory) 
 is applicable,
 when these ratios are small, $l_1,\,l_2,\,R_c /W \ll 1 $, thereby
  $|\delta \varrho_{xy} |/\varrho_{xy} ^{st} \ll 1$.
These inequalities guarantee that the non-hydrodynamic part
of the electron distribution functions,
  which is proportional to the third and higher harmonics by the velocity angle,
  are small as compared with the hydrodynamic part, which consists of the first
 and the second harmonics.

When these ratios~$ l_1,\,l_2,\,R_c /W $ and~$|\delta \varrho_{xy} |/\varrho_{xy} ^{st} $
becomes  comparable to unity, non-hydrodynamic contribution
 in the electron distribution function
becomes important, the flow becomes partly ballistic, and the obtained here results
  for~$\varrho_{xx,xy}$ can substantially change.  Nevertheless we believe that
our theory at the border of  its applicability, when $l_{1,2,0} ,\,R_c\sim W$,
still leads to reasonable and  qualitatively correct results.
 Herewith the smooth inequality,   $|\delta \varrho_{xy} |/\varrho_{xy} ^{st} \lesssim 1$,
for the calculated flows apparently serves as a criterion of
  a qualitative applicability of the current hydrodynamic-like model.
%
%
%
%\\
%\\
\section{3. Temperature dependencies of relaxation times and
corresponding temperature-dependent magnetoresistance }
In this section we present the known results on temperature dependencies
of the  relaxation times which are actual for the developed hydrodynamic-memory
 model of magnetotransport of 2D electrons in high-quality structures.
 First of all,   we imply the high-quality GaAs/AlGaAs quantum wells,
for which the most part of the discussed experiments we performed.

 We assume that the dependencies of the relaxation times
$\tau_0(T)$ and $\tau_2(T)$ on temperature is determined  by a raise 
of electron-electron  collisions rate with temperature,  while the  dependence of the momentum relaxation times~$\tau_1(T)$ 
and~$\tau_1^c(T)$ of  t- and c-electrons on  temperature is due to the  
electron scattering on acoustic phonons.

Indeed, for 2D electron with a quadratic energy spectrum,
 $\varepsilon _ {\mathbf{p}} = p^2/(2m) $,
  that takes place for GaAs/AlGaAs quantum wells, the electron momentum
is proportional to the electron velocity
and momentum is conserved in electron-electron
collisions. Therefore the electron-electron scattering
do not affect the momentum relaxation rates~$  1/\tau_1 (T)   $
and~$    1/\tau_1^c (T) $ of the t- and c-electrons, respectively.
Herewith we neglect the contribution from the electron scattering
on the weak delocalized disorder (the ``dust'') as compared the contribution from
 the scattering on the localized defects, for  the rates in which they
are additive.

Based on these assumptions and taking into account the
presence of temperature-independent disorder-induced contributions in all the
relaxation times,  for the total departure rate of t-electrons
we have~\cite{Qiun,Novikov,Alekseev_Dmitriev}:
\begin{equation}
     \frac{1}{\tau_0(T)} =\frac{1}{\tau_0}
      +  \frac{1}{\tau_{0,ee}}
      \:, \quad
\frac{1}{\tau_{0,ee}}
      =
      \frac{2 \pi }{3 } \frac{
 \ln(\varepsilon _ F / T )
            }{ \hbar \varepsilon _ F }\:T^2
            \:,
            \label{tau_0}
 \end{equation}
 Here $ 1/\tau_0 $ is the departure rate due to scattering on the  localized defects
 and~$1/ \tau_{0,ee}$  is the electron-electron departure rate
within the approximation of weak inter-particle interaction, $r_s \ll 1 $.
 For a strongly interacting 2D electron Fermi liquid, when $r_s  \gtrsim 1 $,
this formula gets some numeric factor,
 depending on  $r_s   $ via the Landau parameters~\cite{Novikov}.
 Formula~(\ref{tau_0})  enters the probability~$P$~(\ref{P})
to make a full cyclotron  rotation for any electron in magnetic field.

The following formula takes place for the relaxation time of the electron
shear stress  (that is, of the viscous part of the momentum flux~$  \Pi_{ij}$)
within the approximation of weak inter-particle interaction,
 $r_s \ll1 $ and in the diapason $T/ \varepsilon _ F \ll r_s$~\cite{Alekseev_Dmitriev}:
\begin{equation}
 \label{tau_2}
     \frac{1}{\tau_2(T)}
     =
     \frac{1}{\tau_2} +
        \frac{1}{\tau_{2,ee}}
      \:, \quad
      \frac{1}{\tau_{2,ee}}
      =
      \frac{8 \pi }{3 } \frac{
 r_s^2 \, \ln(1  / r_s )
            }{ \hbar \varepsilon _ F } \:T^2
            \:,
 \end{equation}
where $1/ \tau_2$ is the rate of  relaxation of the electron momentum flux
due to  scattering on the localized defects. As for Eq.~(\ref{tau_0}),
formula~(\ref{tau_2}) refers  the to t-electrons.

The momentum relaxation time of the t-electrons
 has the following form~\cite{Karpus}:
\begin{equation}
     \frac{1}{\tau_1(T)} = \frac{1}{\tau_1} +
        \frac{1}{\tau_{1,ph}}
      \:, \quad
      \frac{1}{\tau_{1,ph}}
      =
      C \frac{ m E_D ^2 }{ \hbar^3 \rho s^2 a   }
       \:T
       \:.
       \label{tau_1}
 \end{equation}
 This formula is applicable at not too low temperatures,
  $T \gg \sqrt {m s_{l}^2 E_1 } $ (see details in Ref.~~\cite{Karpus}).
Here  $1/ \tau_1$ is the momentum relaxation rate due to scattering
of 2D electrons  on the localized defects; $1/ \tau_{1,ph}$ is
the momentum relaxation rate due to  the electron-phonon scattering;
$C$ is numerical constant that depends on specific geometry
of the quantum well potential;
$E_D$ is the magnitude of the deformation potential controlling
the amplitude of the electron-phonon
interaction; $s_{l}$ is the velocity of the longitudinal
sound, $\rho$~is the  density of the GaAs crystal; $a$~is
the effective width of the quantum well, which determines
the form-factor of the electron-phonon interaction; $E_1 \sim \hbar^2 \pi^2 /(ma^2)$
 is the energy of the level of size quantization of 2D electrons
in the quantum well.

The formula for the momentum relaxation time of the c-electrons  
is similar  to Eq.~(\ref{tau_1}):
\begin{equation}
     \frac{1}{\tau_{1,c}(T)} = \frac{1}{\tau_1^d} +
        \frac{1}{\tau_{1,ph} (T) }
       \:.
       \label{tau_1_c}
\end{equation}
Here $1/ \tau_1^d$ is the rate of relaxation of 2D electron momentum  due to
scattering on the dust, and the electron-phonon scattering rate~$1/ \tau_{1,ph} (T) $
 is  the same as in Eq.~(\ref{tau_1}).
We remind that the residual  scattering  of c-electrons at~$T=0$ is much weaker,
than the scattering of t-electrons on the localized defects: $1/\tau_1^d \ll 1/\tau_1$.

Note that the last two formulas for the momentum relaxation rates
of t- and c-electrons   differ significantly at low temperatures
(due to the last inequality)  and approach each other at very high
temperatures, because of a dominance of the scattering on phonons
in the momentum relaxation.

Using Eqs.~(\ref{tau_0})-(\ref{tau_1_c}), we calculate
 the total  values of the relaxation times~$\tau_{0,1,2}$ and~$\tau_{1}^d $
at different temperatures in realistic samples.
This opens the possibility to find  the mean sample resistivity~$\varrho_{xx} (B,T)$
 of an electron flow as a function of magnetic field~$B$ and temperature~$T$
 in Hall bar samples, examined in various experiments.
Below in this Supplemental material
  we  take the particular parameters of the  2D electron systems,
 those approximately correspond to the moderately-pure samples examined in
the experiments~\cite{Gusev_1,Gusev2} and, particularly, in~\cite{exps_neg_4}. We choose these samples
 because (i) the magnetoresistance curves measure on
 them have relative simple, single-component and smooth, shapes;
 (ii) the 2D electron densities~$n_0$ in these samples are highest 
 and correspond 
  to the lowest interaction parameters~$r_s$.

\begin{figure}[t!]
    \centering
    \includegraphics[width=1.0\linewidth]{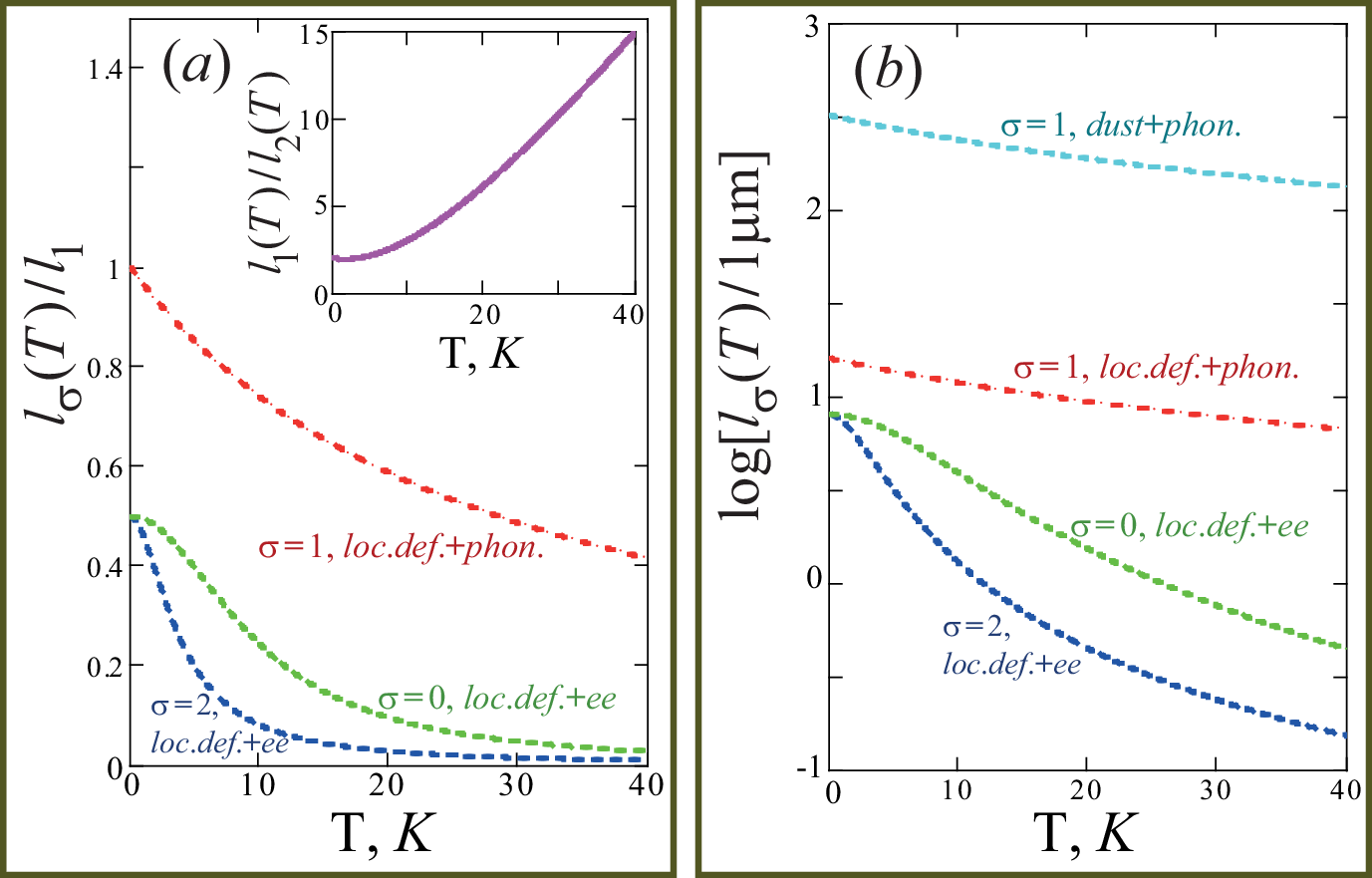}
    \caption{ Temperature dependencies of the relaxation lengths.
    The index $\sigma$  describes the two properties of the relaxation process:
       (i) type of the value whose relaxation is described:
    ``0'' refers to the departure scattering rate,
    that is to relaxation of the whole electron state;
     ``1'' refers to the momentum relaxation;
     ``2'' corresponds to the relaxation of  the viscous momentum flux;
      (ii) the type of the relaxation mechanism:
      ``loc. def.'' refers to the electron scattering on the localized defects;
      ``dust''  denotes the electron scattering
      on the weak disorder between the localized defects;
       ``ee'' means  the electron-electron scattering,
       ``phon'' refers to the electron-phonon scattering
       by the deformation potential mechanism.
       Inset: Ratio of the relaxation lengths
       of momentum and momentum flux as a  function of temperature.}
    \label{pic:relaxation_lengths}
\end{figure}

 Namely, the following values of all the parameters were used below:
($A$)~the characteristics of 2D electrons and their interaction:
the concentration of 2D electrons $n_0=9 \cdot 10^{11} $~cm$^{-2}$,
the effective mass $m = 0.067 m_0$, the 2D Bohr radius $ a_B = 10$~nm,
  the corresponding Fermi energy $\varepsilon_F = 32$~meV,
   the Fermi velocity $v_F = 4.1 \cdot 10^{7}$~cm$/$s, and   
 the interaction parameter  $ r_s = 0.6$,
the dielectric permittivity of the GaAs structure $ \epsilon = 13$;
($B$)~the  characteristics of the electron scattering
 on disorder: $\tau_0/\tau_1 = 1/2$,
$\tau_2/\tau_1 = 1/2$, $ \tau_{d}/\tau_1 = 20 $,
the  scattering length on the localized defects $l_1 = v_F\tau_1 = 16 \: \mu$m
[this values correspond  to the following residual electron mobility at $B=0$
and~$T=0$:  $\mu _0 = 10^6$~cm$^2/$V$\cdot$s];
($C$)~the characteristics of the electron-phonon scattering:
the deformation potential for the interaction
 with longitudinal acoustic phonons $E_D = 7 $~eV,
the speed of sound $s = 5 \cdot 10^5 $~cm$/$s,
 the density of GaAs crystal $\rho = 5.3$~g$/$cm$^{-3}$,
the effective quantum well width $a = 6$~nm;
($D$) the effective  sample width: $W/l_1 =W_{eff}/l_1 = 1$.
The last two values are of the order of their nominal values
  in experiments~\cite{Gusev_1,Gusev2,exps_neg_4}     and qualitatively account
the complex profile of  the potential the quantum well and 
the complex shape of
the sample edges as well as a non-zero slipping of the electron flow 
to the edges (see also analysis of data from Ref.~\cite{exps_neg_4} in Ref.~\cite{je_visc}).

\begin{figure*}[t!]
\centerline{\includegraphics[width=0.75 \linewidth]{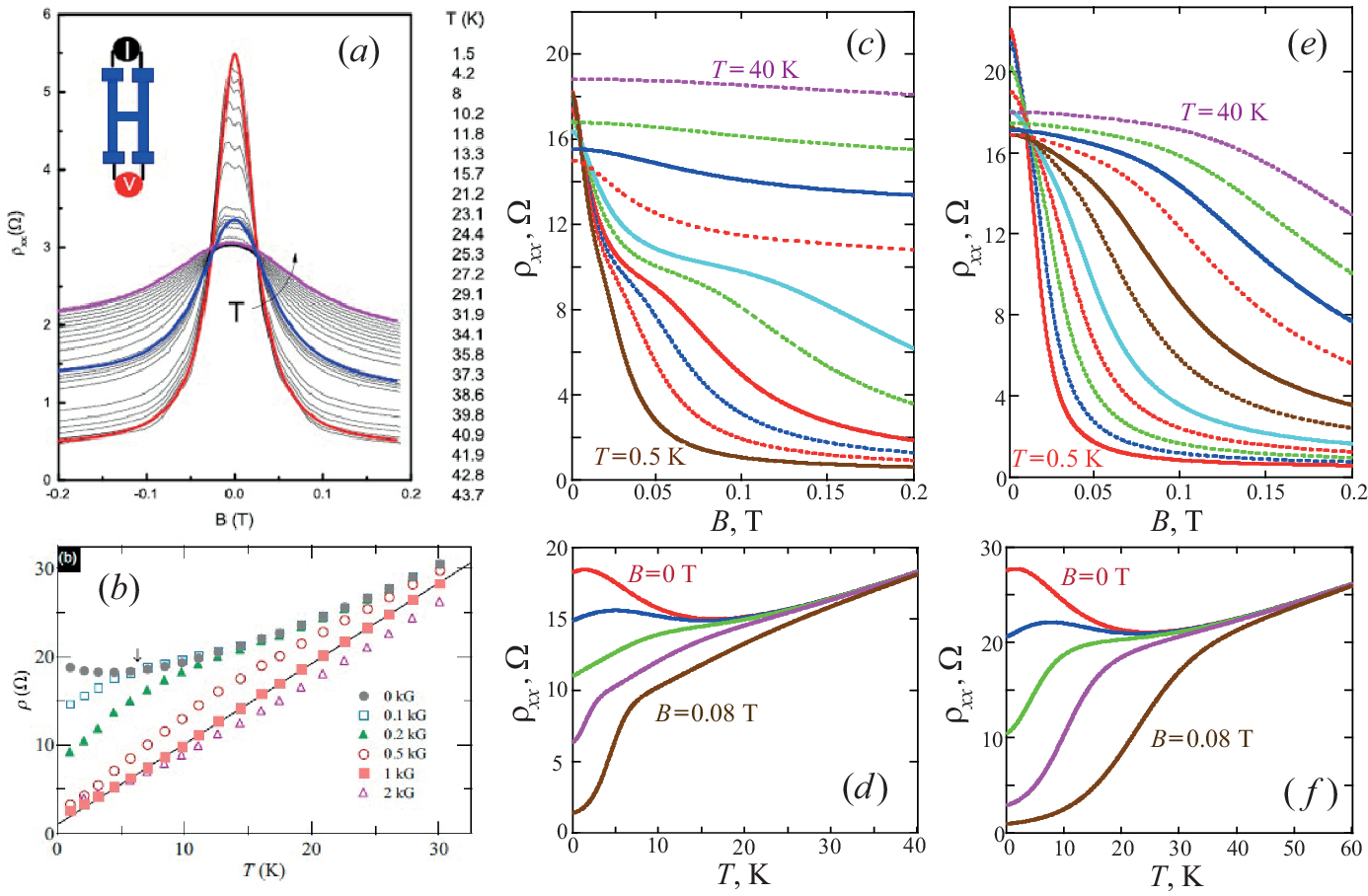}}
\caption{
  ($a$):~Dependencies of the sample mean resistivity~$\varrho_{xx}$
  on magnetic field~$B$
at different temperatures~$T$ obtained in  experiment~\cite{Gusev_1}
for ultra-pure samples of GaAs quantum wells.
($b$):~Dependencies of the sample mean resistivity~$\varrho_{xx}$
on temperature~$T$
at various magnetic fields~$B$  measured in Ref.~\cite{exps_neg_4}.
($c$-$d$):~Results of
our theory for $\varrho_{xx}(B,T)$. The parameters taken
for calculation are similar to the ones
 of the structures examined in Refs.~\cite{Gusev_1,exps_neg_4}, in particular:
 $n_0 = 9 \cdot 10^{11} $cm$^{-2}$,
  $ \tau_0/ \tau_1 = 0.5$, $ \tau_2/ \tau_1 = 0.5$, $ \tau_1^d/ \tau_1 = 20 $,
$l_1 = 16\; \mu$m, the effective width
of the quantum well~$a = 6$~nm, the effective  sample width~$W/l_1 = 1$,
 others parameters are described in  the text in this section [points~($A$)-($D$)].
($e$-$f$):~The same as in panels ($c$-$d$) for the shifted parameters of the 2D system:
    $ \tau_0'/ \tau_1 = 1$, $W'/l'_1 = 0.9$,  $l_1' = 12\; \mu$m,
    $1/\tau_{0,ee}' (T) \, = \,  0.1\, [1/\tau_{0,ee} (T)] $,
 where the dependence~$\tau_{0,ee} (T)$, the parameters
 $\tau_1$,   $l_1$, and
 the all other parameters  are   the same as for panels~($c$-$d$).
  }
\end{figure*}

In Fig.~S3 we present  the temperature dependencies of all the  relaxation lengths
those correspond to the relaxation rates presented above:
$l_\sigma(T) = v_F \tau_\sigma(T)$.
The index $\sigma$  here explicitly describes
 the two properties of the relaxation process:
the  type of the value whose relaxation is described and
 the type of the relaxation mechanism.
Note that $l_0(T) =  l_{0,\,loc.def.+ee}(T)$ and $l_2(T)=l_{2,\,loc.def.+ee}(T) $  decrease
faster with the increase of temperature than both $l_{1} (T) =l_{1, \,loc.def.+phon.} (T) $
and~$l_{1,c} =l_{1, \,dust +phon.} (T) $.  This happens because the electron-electron
collisions rate grows faster with temperature, as~$T^2$,  than the rate of
the electron-phonon  scattering, growing as~$T$ (and, obviously, due to the particular
values of the coefficients  in these two dependencies for the GaAs quantum wells).

Next,  we substitute the presented  dependencies~$\tau_\sigma (T)$ and~$l_\sigma (T)$,
Eqs.~(\ref{tau_0})-(\ref{tau_1_c}),
  into Eq.~(\ref{res_I})   and obtain the temperature- and magnetic-field-dependent
mean sample resistivity~$\varrho_{xx}(B,T)$.

In  Fig.~S4 we cite the experimental data on resistivity~$\varrho_{xx}(B,T)$
obtained in Refs.~\cite{Gusev_1,exps_neg_4}
  for the Hall bar samples of GaAs quantum wells
  with not too high mobilities of electrons [panels~($a,b$)]
  and  present the results of our calculations of~$\varrho_{xx}(B,T)$  [panels~($c,d,e,f$)].
In Figs.~S4($a,c,e$) are presented the resistivity~$\varrho_{xx}(B)$
as a function of magnetic field~$B$
at different temperatures~$T$,  while in Figs.~S4($b,d,f$)
we present temperature dependence
of the resistivity~$\varrho_{xx} (T)$ at various magnetic fields~$B$.
From the upper panels we can trace the evolution
of the giant negative magnetoresistance  with temperature.
In the lower panels we  see the decrease of resistance with temperature
at lower magnetic fields (the Gurzhi effect)
 at and its increase at  stronger magnetic fields.

In panels~($c,d$) we show the results for~$\varrho_{xx}(B,T)$ 
at  the relaxation
rates~(\ref{tau_0})-(\ref{tau_1_c})
 with the literal values of the above presented parameters~($A$)-($D$),
  while  in  panels~($e,f$) we plot the functions~$\varrho_{xx}(B,T)$ for
a shifted set of the above parameters.
 The last calculation is done  in order to demonstrate
 various possible types of the dependencies~$\varrho_{xx}(B,T)$,
those can be obtained within the developed theory.
These shifts of parameters can qualitatively account
  for possible  complex properties  of the real systems  as compared to
 those are  supposed in our simple model.  Such more complex properties may be:
(i) more complex types of  the memory effects [in particular,
  due to the  smooth profile of the potential~$U_C^{scr} (r)$
   of localized defects, see Fig.~S1($a$)];
(ii) a relatively strong inter-particle interaction
[we have $ r_s = 0.6$ for the above density~$n_0 = 9 \cdot 10^{11}$~cm$^{-2}$,
which is still on the border of the applicability the Fermi gas model
  of Ref.~\cite{Alekseev_Dmitriev} used
  for calculation of~$1/\tau_{0,ee}$~(\ref{tau_0}) 
  and~$1/\tau_{2,ee}$~(\ref{tau_2})];
(iii) more complex shapes of electron flow
[due to a bent shape of the sample edges
   and macroscopic  inhomogeneities  inside the sample].

The main difference between the curves in panels~($c$)    and~($e$) is as follows.
For the panel~($c$) [and~($d$)], the times~$\tau_2(T)$ and $1/\tau_0(T)$ differ substantially,
 therefore the curves $\varrho_{xx}(B)$ consist of the two distinct parts
with the width of the order of~$1/[2\tau_2(T)]$ (at lower~$B$)
 and~$2\pi/\tau_0(T)$ (at higher~$B$), the first of    which is controlled
mainly by  the viscosity effect,   while the second of which is induced
by the memory effects.  For the panel~($e$) [and~($f$)],
the values~$1/[2\tau_0(T)]$ and~$2\pi/\tau_2(T)$ are chosen to be close
 one to another,
 therefore the curves~$\varrho_{xx}(B)$  have a single-component shapes and
 are simultaneously controlled  by  both the viscosity and the memory effects
 in comparable degrees.

Furthermore, it can be seen from Fig.~S4($c$) that,
at the first presented above set of the parameters~($A$)-($D$),   the shape of
 the resistivity~$\varrho_{xx}(B)$ at relatively low~$B$ (where $\beta \lesssim 1 $)
is mainly corresponds to the shape of the Ohmic-hydrodynamic Gurzhi   contribution:
\begin{equation}
  \varrho_{xx}  ^{G} (B) \, = \, 
  \frac{ \varrho_D  } { 1-\tanh \xi /\xi}
  \:,
\end{equation}
 where now $ \varrho_D  = m/[n_0 e^2 \tau _{1} (T)]$
 and $\xi = \xi(T)$ also depends on~$T$ via $\tau _{1} (T)$~(\ref{tau_1}).
    In the region of higher~$B$, where $\beta \gtrsim 2 \pi $
the magnetotransport is predominantly Ohmic with the memory effects,  therefore
 $\varrho_{xx}(B)$ behaves according Eq.~(\ref{MR_ohm_mem_narr}).
  Except the diapason of low~$B$ and~$T$,  the increase in the rate of
phonon scattering with~$T$ leads to the growth of  the resistance
at given~$B$, being linear in~$T$ at sufficiently high~$T$ [see Eq.~(\ref{tau_1})].
 For the shifted set of the parameters (see the caption of Fig.~S4),
 both the hydrodynamic and the Ohmic-memory contributions provide comparable contributions to
the function~$\varrho_{xx}(B, T )$  in the whole diapason of~$B$ and~$T$ [Figs.~S4($e$)]. As a result,
  the magnetoresistance curves in panels~($c$) and~($e$) have substantially different
shapes: two-component and single-component, respectively.
 The behavior of the dependencies $\varrho_{xx} (T)$ at different~$B$
 in Figs.~4($d,f$) reflects the  dependencies~$ \varrho_{xx} (B)$  in Figs.~4($c,e$)  via another way.

It is seen from Figs.~S4($c,e$) that
the width of the magnetoresistance curves
 increases monotonically with~$T$. This occurs both due to the growth
of the shear stress relaxation rate~$1/\tau_2 (T)$
and the departure rate~$1/\tau_0 (T)$. The first one
 determines the width of
 the hydrodynamic-dominated contribution to the curve~$\varrho_{xx}(B)$
 at $\beta \lesssim 1$ in panel~($c$).
The second one  controls the width of
the Ohmic-memory  contribution to~$\varrho_{xx}(B)$,
dominating at $\beta \gtrsim 2\pi$ in panel~($c$).
As we said above, for the curves in panel~($e$)
both the rates ~$1/\tau_2 (T)$ and~$1/\tau_0 (T)$
 are comparable and, thus, together determine the widths
 of the single-component smooth curves~$\varrho_{xx}(B)$.

The experimental curves in Fig.~S4($a$) are more similar
to the results of our calculation Fig.~S4($e$) for the second, shifted, set of parameters (presented inside the figure caption). However, there are many experiments in which
the shape of the magnetoresistance curves is two-component and is more similar
to the curves shown in Fig.~S4($c$) for the first set of parameters~($A$)-($D$)
 (see also Fig.~2 in the main text and Fig.~S4 in the next section).

The temperature dependencies of the experimental resistivity in Fig.~S4($b$)
 are  similar 
  to the results of  our calculations for both sets of parameters
 at   sufficiently low magnetic fields and Temperatures
and differ significantly at their larger values  [compare panels~($b$) and~($d,f$)].
 Possibly, the reason for such relation between the experiment and the current theory  is as follows. The rates $\Gamma_{t,c}$ of the $ t \leftrightarrow c $ transitions, entering Eq.~(\ref{bal_eq_dens}), growth rather fast with
 the increase of temperature (primary, due to the ee-scattering) as well as of  magnetic field (possibly, strong magnetic fields can
affect  the very scattering processes). The resulting  increase 
 of~$\Gamma_{t,c}$ leads to mixing  of t- and c-electrons,
  violation of the condition $W \lesssim l_{tc}(T)$, and 
 a suppress of  the memory effects.
  Indeed, an almost linear dependence, $\varrho_{xx}(T) \sim T$,
 is observed experimentally at sufficiently large~$B$ and~$T$,
  that, apparently, corresponds to the dominance 
the Ohmic contribution to the resistance related to the electron-phonon scattering:
$\varrho_{xx} \sim m/[n_0 e^2 \tau_{1,ph}(T)]$.
%
%  % ! here 
%
%
%
% \\
% \\
\section{4. Review of experimental data on  classical magnetotransport
in high-quality samples }
Let us  review the experimental data on  magnetotransport
 of 2D electron in high-quality nanostructures
in classical regime (that is, in magnetic fields below
 the quantizing regime).  Such review is instructive for a general understanding
of the relevance of the   magnetoresistance
 following from our model to the experimental data on
the giant negative magnetoresistance in various high-quality samples.

In Fig.~S5 we present some available experimental data on magnetoresistance
  of high-quality GaAs quantum well and graphene samples at varying relevant parameters
of the examined  systems: temperature, the in-plane component of magnetic field,
the 2D electron densities.

The giant negative magnetoresistance in all cited  experiments
exhibits a number of common features.

The main of these features, obviously,  is a strong drop in resistance with the increase
of  magnetic field.  This drop occurs in fields near some characteristic $B^*$  field,
varying in the range from~0.2~kG up to 1~kG.
The other of these features  are  the  strong dependencies of    the magnetoresistance
on temperature and the in-plane component of  magnetic field,    namely, more or less
fast suppression of the giant negative  magnetoresistance with increasing these  two values.

Furthermore, the giant negative magnetoresistance exhibits certain   more special
features,  which  crucially depend  on the sample type.

The relative magnitude of the resistance drop,
 $ r = \varrho_{xx}(B=0)/\varrho_{xx}(B \gg B*$), varies from sample to sample
 in the diapason from  the values of the order of unity, $r \sim 1$,   
  [Figs.~S5($a,d,i$)]
up to the values about~$ r \sim 50 $   [Figs.~S5($e,g,i,j,k,l,p$)].
 The absolute value of the mean sample resistivity
in the region of lowest magnetic fields, $B \to 0 $, also varies from sample to sample:
from the values  10-20~Ohms [Figs.~S5($e,g,l$)]
to the values   2-0.5~Ohms [Figs.~S5($b,h,i,p$)].

The shape of the main part of the magnetoresistance curves at the lowest temperatures
 is also can be different: in some samples, it is quite blunt, resembling
  the roof of a house [Figs.~S5($a$)-($d$)],
while in others, this curve consists primarily of a single sharp peak [Figs.~S5($e,g,h$)].
In some samples, as it has been already discussed in the main text [see Fig.~2($c,d$)],
the curve consists of the two more or less distinct parts [Figs.~S5($i$)-($l$)]:
 a narrow, small or not so small peak and a wider main part of the curve,
decreasing with magnetic field.
Herewith the narrow peak near zero magnetic field can be dependent [Fig.~S5($j$)]
or independent [Figs.~S5($a$)-($c$)] on temperature (in the diapason of low
temperatures).

\begin{figure*}[th!]
    \includegraphics[width=0.99\textwidth]{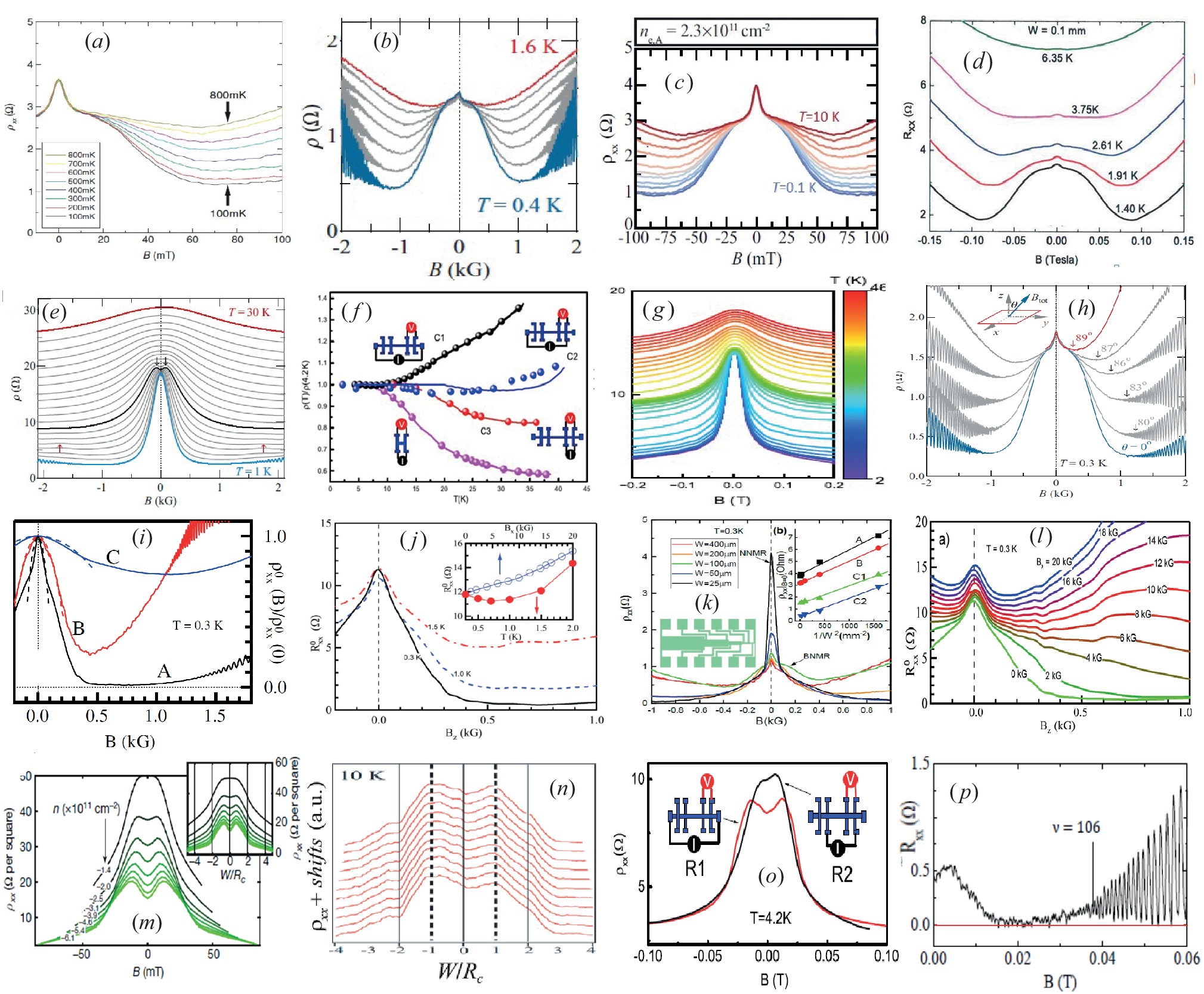}
    \caption{
Experimental data on electric transport in  high-quality structures
 with 2D electrons,  obtained   by different experimental groups.
  Panels ($a$)-($p$) cite the data from the following works:
($a$) from Ref.~\cite{exps_neg_3};    ($b$)~-~\cite{exps_neg_1};
                  ($c$)~-~\cite{d1_new}; ($d$)~-~\cite{exps_neg_2};
($e$)~-~\cite{exps_neg_4}; ($f$)~-~\cite{Gusev_1};
              ($g$)~-~\cite{Levin2024}; ($h$)~-~\cite{exps_neg_1};
($i$)~-~\cite{Dai2010};  ($j$)~-~\cite{Dai_Stone};
                ($k$)~-~\cite{recentest_};   ($l$)~-~\cite{Dai_Stone};
($m$)~-~\cite{Nature};   ($n$)~-~\cite{Nature2};
             ($o$)~-~\cite{Gusev2}; ($p$)~-~\cite{Nature_Materials}.
 \\
{\color{white} $.\:\;\:$~}All panels, except panels~($m$),($n$), present the data
for high-quality GaAs quantum wells;
 in panels~($m$),($n$) magnetoresistance of  graphene stripes is presented.
Mean sample longitudinal resistivity~$\varrho_{xx}$
as a function of magnetic field~$B$
 is presented in all panels, except panel~($f$),
in which the resistance as a function of temperature~$T$
 at zero magnetic field is shown.
  \\
{\color{white} $.\:\;\:$~}On each panel, different curves correspond: to
gradually changing temperatures~$T$ [($a$)-($e$),($g$),($j$)];
to different samples with higher and lower defect densities~($i$);
to different  sections, with various widths~$W$, of a given sample~($k$),
to various 2D electron densities~$n_0$~($m$),($n$);
to gradually changing  in-plane component of magnetic field~($h$),($l$);
to different arrangements of contacts to the samples~($f$),($o$).
 \\
{\color{white} $.\:\;\:$~}In inset of panel~($j$)
the red line  is the resistivity~$\varrho_{xx}$
at zero perpendicular component of magnetic field, $B \equiv B_z=0$,
as a  function of temperature, while
the blue line is the resistivity~$\varrho_{xx}$ at~$ B_z= 0$
 as a  function of  in-plane component~$B_x$.
   \\
{\color{white} $.\; \: \,  $~}In inset of panel~($k$) is shown the resistivity
at $B=0$ as function of  the inverse squared sample width~$1/W^2$, that demonstrates
 a large contribution   of the hydrodynamic  dependence,~$\varrho_{xx} \sim 1/W^2$,
in the observed resistivity.
\\
{\color{white} $.\:\:\; $~}In panel~($n$) and in  inset of panel~($m$)
the resistivity as function
of magnetic field in dimensional units,  $W/R_c = (W /v_F)\, \omega_c $, 
 are plotted  for various 2D electron densities.
       }
\end{figure*}

The specific character of the evolution of
the giant negative magnetoresistance with temperature also varies from sample to sample.
With increasing temperature, either a broadening and suppression of the amplitude
of the negative magnetoresistance occur [Figs.~S5($e,g,j$)]
or only a gradual decrease in amplitude takes place [Figs.~S5($a$)-($d$)].
 In some samples, resistance decreases with temperature at zero and low magnetic fields,
that is, the Gurzhi effect is observed
[see Figs.~S4($a,b$), Fig.~S5($f$), and inset in Fig.~S5($j$)].

In some samples, the resistance~$\varrho_{xx} (B) $  at low temperatures has
 a maximum at a low magnetic field,~$B_0 \ll B^*$,
 and a dip at   zero magnetic field,~$B = 0 $,
resulting in a curve resembling horns [Figs.~S5($m$)-($o$)].
In some samples, in a similar field range  $-B_0<B<B_0$, where $ B_0 \ll B^*$,
a very small peak appears [Figs.~S5($a$)-($d$),($h$),($l$)], which is,
at sufficiently low temperatures, independent of temperature [Figs.~S5($a$)-($d$)]
and of the in-plane component of the magnetic field [Figs.~S5($h,l$)].
Moreover, in samples with a small temperature-independent peak,
 the main temperature-dependent part of the magnetoresistance curve often
has  a characteristic blunt roof-like shape [Figs.~S5($a$)-($d$)].

Now let us comment how the predictions of our theory relate
to the just reviewed various types and features
of the observed giant negative magnetoresistance.

We remind that the developed theory has the two sources of negative magnetoresistance:
 (i)~the memory effects at scattering on the localized defects in magnetic field
(in the absence of transitions between the c- and t-components)
and (ii)~the formation of spatially inhomogeneous hydrodynamic flows
due to  unconventional  viscosity induced by the scattering on defects in bulk and rough edges of the sample.
 Both of these contributions are accounted in formula~(\ref{res_I}).

If the memory-induced contribution (i) dominates (apparently, this
 is the case for sufficiently wide samples,
$ l_{1,2} \ll W \lesssim l_{tc} $), then
the magnetoresistance takes the form (\ref{MR_ohm_mem_narr}).
It has a blunt profile and disappears at sufficiently low temperatures,
 due to the fairly strong temperature dependence of $\tau_0(T)$~(\ref{tau_0}).
 The magnetoresistance curves with very similar shape
 were observed, for example, in works whose results are presented in Figs.~S5($a$)-($d$)
 [compare them with Fig.~S2($a$)].
With the increase of temperature such negative magnetoresistance disappears
 in experiments rather fast,
that is  explained by the increase of the departure rate $1/ \tau _0(T) $
 according to Eqs.~(\ref{tau_0}) and the corresponding decrease of
 the probability~$P= e^{-2 \pi /[\omega_c \tau _0(T)] }$
controlling the magnitude  of the memory effects.
Herewith some additional mechanisms apparently contribute to transport
 in such samples, leading to the appearance of a positive magnetoresistance 
 also  [see Figs.~S5($a$)-($d$)]. Possibly, the last effect can be induced 
by the change of the regimes~$W\lesssim l_{tc} (B)$ 
and~$W \gg l_{tc} (B)$.

The  hydrodynamic-like  contribution (ii)  [described by the term with $\xi$
 in formula~(\ref{res_I})] can dominate
  in sufficiently narrow samples: $W \ll l_{tc} $,
 $W \sim l_{1,2}$. In this case
 the magnetoresistance has a characteristic sharp
 Lorentzian-like  profile at low temperatures. Such type magnetoresistance 
 was
  observed in the experiments  whose data are cited  in Fig.~S4($a$) and Figs.~S5($e,g$).
With increasing temperature, its  profile gradually broadens 
[Figs.~S4($a$) and~S5($e,g$)].
 and the absolute value of the resistance gradually increases
 in all [Figs.~S5($e,g$)] or only in sufficiently high [Fig.~S4($a$)] 
  magnetic fields.
These broadening and increase  are explained, respectively,
by the contributions
 to the momentum flux relaxation rate~$1/\tau_2(T)$
and to than the departure rate~$1/ \tau _0(T) $ from  electron-electron collisions
[Eqs.~(\ref{tau_0}) and~(\ref{tau_2})] and
by the contribution to the momentum relaxation rate~$1/\tau_1(T)$
from the electron-phonon scattering [Eq.~(\ref{tau_1})].

Formula (\ref{res_I}) also leads to the coexistence of the two types
of magnetoresistance described above. This takes place at the comparable,
but different rates~$1/\tau_2$ and~$1/\tau_0$, which control the widths
of the hydrodynamic and the memory-induced
contributions to the dependence~$\varrho_{xx} (B)$, at
the rather small sample widths, $W \sim l_1$.
 Such two-component magnetoresistance   was apparently observed
 in the experiments whose results are presented  in Figs.~S5($i$)-($l$).

The temperature-independent small peak  [see Figs.~S5($a$)-($d$)]
and the dip  [see Figs.~S5($m$)-($o$)]
in the region of very low magnetic fields, $|B| < B_0 \ll B^*$
have likely the ballistic nature~\cite{Scaffidi2017,a,b}
or, possibly, this  the small peak is due to the memory effects 
 at scattering of 2D electrons on big, macroscopic,
 defects~\cite{d1}, which are present  in some samples.
The main evidence of the  ballistic or macroscopic-defect-induced 
 nature of the small peak,
along with its independence of temperature and in-plane magnetic field,
 is that it exists in the  magnetic field range, $|B| < B_0 $,
  corresponding to cyclotron diameters
of unperturbed electron trajectories~$2R_c$ equal and larger
 than the sample width $W$ (for moderately wide  samples)
 or the size~$W_{eff}$
 of macroscopic inhomogeneities  in  very wide samples
[such values~$2R_c(B_0)$ are   of the order of  50~$\mu$m].

Let us pay special attention to the experimental data  from Ref.~\cite{Nature_Materials},
presented in Fig.~S5($p$).

In this experiment, the samples of record quality (with low-temperature
mobilities~$\sim 40 \cdot 10^ {6}$~cm$^2/$V$\cdot$s) and large size (several millimeters).
The measured magnetoresistance  amplitude~$\varrho_{xx} (B=0)/ \varrho_{xx} (B=0)$
is very large, of the order of 20, and the magnetic field $B^*  \approx  10 $~mT,
at which  the drop of the curve $\varrho_{xx} (B)$ occurs,
is especially low  as compared to the other curves in Fig.~S5.
The absolute value of the measured mean sample resistivity is also extremely low,
0.5~Ohm at~$B=0$.  Furthermore, as can be seen from  Fig.~S5($p$),
  fluctuations in the measured resistance are very large.  It is also seen
that quantum Shubnikov  oscillations in  $\varrho_{xx}(B)$  begin in very  weak
magnetic fields~$B \approx   40 $~mT=0.4~kG.  All these features are, apparently, the signs of
an extremely high quality of the examined  sample.

However, the behavior of the curve  in Fig.~S5($p$) in the region
 of very low magnetic fields is rather unclear. Specifically, it is unclear whether:
(i) the maximum at the  low magnetic field
$B_0 \approx  5 $~mT$< B^*$ is, in fact, the  maximum
 at zero magnetic field~$B=0$ in the  diagonal resistivity~$\varrho_{xx} (B)$,
being shifted due to the admixture of the Hall resistivity~$\varrho_{xy} (B)$
to the measured signal  due to the imperfect geometry of the contacts;
or (ii)  this maximum at~$B_0$ is the proper maximum
in the function~$\varrho_{xx} (B)$, located at such a non-zero magnetic field,
 while there  is a dip in~$\varrho_{xx} (B)$
at zero field~$B=0$ [that is the ``horns''  appear  in the curve~$\varrho_{xx} (B)$, as
in Figs.~S5($m$)-($o$)].

In any case, this sample exhibits a very large magnetoresistance amplitude
$r=\varrho_{xx} (B=0)/ \varrho_{xx} (B=0)$,
a record-small absolute value of the resistivity~$\varrho_{xx} (B=0)$,
 and a very small characteristic field~$B^*$, where $\varrho_{xx} (B )$ drops significantly.
Within the framework of our theory, such values corresponds to extremely  weak relaxation 
rates~$1/\tau_{1,0,2}$ and~$1/\tau_1^d $ of  the electron scattering
on both   the localized defects and the delocalized weaker disorder
between them.
\\
\section{5. Possible developments of proposed hydrodynamic-memory model of magnetotransport }
One next aim within  the   theory  of the magnetotransport
of  2D electrons  due to the memory and the hydrodynamic effects
is the study of  the case of  the wide samples,
$W\gtrsim l_{tc}$, where the transitions $t \leftrightarrow c $
becomes important (in any or sufficiently high magnetic fields).

As we mentioned above, these transitions induce perturbations of the c- and t-electron
densities~$\delta n ^\alpha$  according to Eqs.~(\ref{bal_eq_dens}).
 Our analysis shows that, in wide samples,  the density perturbations~$\delta n ^\alpha $
contribute not only to the Hall effect, but also to  the flows~$ \mathbf{q} ^ \alpha (y)$
and, thus, to the total current~$I$.  As a result, the solution of the equations for~$ \mathbf{q} ^ \alpha (y)$  will lead to
the crossover between the studied here regime  of relatively narrow samples,
$W \ll  l_{tc}$, in which t- and c-electrons do not transform one to another
and the magnetoresistance is given by Eq.~(\ref{MR_ohm_mem_narr}),
 to the wide-sample-regime ($W \gg l_{tc}$), studied in Refs.~\cite{m1}-\cite{m4}. In the last regime the $t \leftrightarrow  c$ transitions do actually occur in
 the neat-edge regions, the bulk contribution in~$I$ dominates,
 and  the magnetoresistance is approximately described
by the formula~$\varrho_{xx}(B) / \varrho_ D \approx [1-P(B)] $~(\ref{BEM}).

 Another regime  which can be studied on base of our theory is
the ballistic-hydrodynamic regime, realized in the  narrower samples,
with the  widths~$W \lesssim l_{1,2,0}$.

% ! here 

  Although in this work we have already performed the calculations of the flows and the resistances~$ \varrho_{xx,xy}$ 
for the  samples with relatively small widths, $W \sim l_1$,
  these results   should be considered as the results of
  a qualitative type of accuracy, since the hydrodynamic approximation
for the distribution function is strictly speaking inapplicable at such~$W$.
 At $W \sim l_1$ it is necessary to take into account non-hydrodynamic contributions in the distribution function, which can provide comparable contributions
 to the flow characteristics.
As  it was mentioned above in Section~2 of SM,  a  characteristic
of the applicability of the current  purely hydrodynamic consideration
 is the smallness of the resulting correction to
the Hall resistance~$\delta \varrho_{xy}$    relative to
its standard value~$\varrho_{xy}^{st}$.     The above calculations
for samples with $  l_2 \lesssim l_1 \sim W $ yield for  $|\delta \varrho_{xy} |/ \varrho_{xy}^{st}$
    the values      of the order of unity, but smaller than unity [see Fig.~S2($b$)].
 So the above calculations for the samples with such~$W $
 are reasonable, but lie on the border of applicability of the theory.

In this way, for narrow samples,~$W \sim l_{1,0,2}$,
 it is of  interest to study
the non-hydrodynamic, ballistic-like, contributions  in the distribution functions
of t- and c-electrons,  being proportional to the third and higher
harmonics by
 the   velocity angle, and the corresponding corrections
  in the flows and the resistivities~$\varrho_{xx,xy} (B)$.
 In small magnetic fields,  when~$W \lesssim R_c $, these contributions
are substantial  in the whole sample, while at higher fields, when $W \gg R_c $
they become important only in the near-edge layers of
 the widths~$\sim R_c$. The semi-ballistic flows in these   regions
 are controlled by the electron scattering on defects as well as
on the sample edges. One can expect that the ballistic contributions 
 in the flows should
lead to a non-hydrodynamic features in
 the dependencies~$\varrho_{xx}(B)$ and~$\varrho_{xy}(B)$
at low magnetic fields, when~$W \lesssim R_c $, which  may provide 
the explanations  of the ``horns''
and small temperature-independent peak in the region $B <B_0$~[see Figs.~S5($a$)-($d$), ($h$), and~($l$)-($o$)].


\begin{thebibliography}{}
\bibitem{f}
It should be noted that the direct use of the Ohmic-hydrodynamic model
of Refs.~\cite{Gurzhi_rev,je_visc} for the quantitative explanation
of experiments \cite{exps_neg_1,exps_neg_2,exps_neg_3,exps_neg_4,Gusev_1} 
(and others)  on the giant negative magnetoresistance
leads to an anomalous, unrealistically large, ratio~$\tau_1/\tau_2$ of
the residual (in the limit~$T \to 0 $) relaxation times  of momentum,~$\tau_1$, and
of shear stress,~$\tau_2$, due to the scattering of 2D electrons
on defects~\cite{je_visc,Gusev_1,Maxim}.
\\
\bibitem{f3}
We remind that, as it often occurs in disordered systems
at low temperatures, the existence of nearly-equilibrium  distributions
of both the t- and c-electrons implies that a very weak inter-particle
scattering  still takes place,  controlling the electron distributions
over energy,  but do not  substantially affect their transport properties.
\\
\bibitem{f4}
A derivation of such balance equations from
the microscopic kinetic equations  (for the different groups of electrons) is
extremely difficult as well as ambiguous, especially in the presence of the two types of disorder. Indeed,  even  a  derivation
in Refs.~\cite{m2,m3} of the Ohmic  transport equation  from the kinetic equation
 for the much simpler case of  spatially uniform flows  in a sample with only
the  localized defects is very cumbersome  and has the points,
which are difficult to prove.
\\
\bibitem{ff}
Indeed, in the very low magnetic fields, 
 $ \omega _c \lesssim  1 / \tau_1^d $,
the density of the c-electrons, 
$n_0P(\omega_c)$ is exponentially small, therefore
their contribution to the total momentum flux, 
$   n_0 P \boldsymbol{\Pi}^c  $,
is much smaller than the one from 
the t-electrons, $  n_0 (1-P) \boldsymbol{\Pi}^t  $.
 In the rest diapason of moderate magnetic fields,
 $ 1 / \tau_1^d \ll \omega _c \lesssim  1/\tau_1 $,
 when the hydrodynamic contribution to the flow of t-electrons is substantial,
 the momentum flux of c-electrons $ n_0 P  \boldsymbol{\Pi}^c  $
and their velocity gradient~$(V_x^c)'$  becomes related via
the viscosity $\eta_{xx} ^d  \, \approx  \, m v_F^2/(16  \, \omega _c^2 \, \tau_2^d )$
due to the scattering on the dust by the formulas, being similar to the relations
(presented in the first paragraph of Section~3) between
 the  values $ n_0 P  \boldsymbol{\Pi}^t $ and~$(V_x^t)'$ for  t-electrons  via
 the viscosity $\eta_{xx} $ due to the scattering  on the localized defects.
The above value  $\eta_{xx} ^d$ in this diapason of~$\omega _c $ is much smaller
than~$\eta_{xx} $ (as $1/\tau_2^d \ll 1/\tau_2 $). Thus, provided the contributions
 to the flow density   from c- and from  t-electrons  are comparable
[this takes place when $ (\tau_1^d /\tau^1) P \sim (1-\tanh\xi/\xi)(1-P) $,
see Eq.~(\ref{res_I})],  then the values $ \boldsymbol{\Pi}^c $ and~$ n_0 P \boldsymbol{\Pi}^c $ become much smaller
than $ \boldsymbol{\Pi}^t $ and~$ n_0(1-P)\boldsymbol{\Pi}^t $.
\end{thebibliography}
\end{document}